\documentclass[pre,twocolumn,amsmath,amssymb,floatfix]{revtex4}

\usepackage{graphicx}
\usepackage{dcolumn}
\usepackage{bm}        
\usepackage[percent]{overpic}
\usepackage{color}

\def\ADDA#1{{#1}}        
\def\ADDB#1{{#1}}        

\begin{document}

\title{Inverse cascades and resonant triads in rotating and stratified
  turbulence}
\author{D. Oks$^{1,2}$, P.D.~Mininni$^{1,3}$, R. Marino$^{4}$, and  
  A. Pouquet$^{3,5}$}
\affiliation{$^1$ Universidad de Buenos Aires, Facultad de Ciencias 
  Exactas y Naturales, Departamento de F\'\i sica, \& IFIBA, CONICET, 
  Ciudad Universitaria, Buenos Aires 1428, Argentina.\\
                  $^2$ Laboratoire de Physique - UMR 5672, 
  Ecole Normale Sup\'erieure de Lyon / CNRS, 46 All\'ee d'Italie,
  69007 Lyon, France\\
                  $^3$ NCAR, P.O. Box 3000, Boulder, Colorado
  80307-3000, USA.\\
                  $^4$ Laboratoire de M\'{e}canique des Fluides et
  d'Acoustique, CNRS, \'{E}cole Centrale de Lyon, Universit\'{e} de
  Lyon, 69134 \'{E}cully, France.\\
                  $^5$ Laboratory for Atmospheric and Space Physics,
  CU, Boulder, Colorado 80309-256, USA.}
\date{\today}

\begin{abstract}
Kraichnan seminal ideas on inverse cascades yielded new tools to study
common phenomena in geophysical turbulent flows. In the atmosphere and
the oceans, rotation and stratification result in a flow that can be
approximated as two-dimensional at very large scales, 
but which requires considering three-dimensional effects to fully
describe turbulent transport processes and non-linear
phenomena. Motions can thus be classified into two classes: fast 
modes consisting of inertia-gravity waves, and slow quasi-geostrophic
modes for which the Coriolis force and horizontal pressure gradients
are close to balance. In this paper we review previous results on the
strength of the inverse cascade in rotating and stratified flows, and
then present new results on the effect of varying the strength of
rotation and stratification (measured by \ADDA{the inverse Prandtl
  ratio $N/f$}, of the Coriolis frequency to the Brunt-V\"ais\"ala
frequency) on the amplitude of the waves and on the flow
quasi-geostrophic behavior. We show that the inverse cascade is 
more efficient in the range of $N/f$ for which resonant triads do not
exist, $1/2 \le N/f \le 2$. We then use the spatio-temporal spectrum
to show that in this range slow modes dominate the dynamics, while the
strength of the waves (and their relevance in the flow dynamics) is
weaker.
\end{abstract}
\maketitle

\section{Introduction}

Kraichnan landmark paper on inverse cascades in two-dimensional (2D)
turbulence \cite{Kraichnan67} has been the stepping stone for a
substantial fraction of the research carried out in geophysical
turbulence in the last 50 years. Besides introducing the concept of a
range of scales in which energy can flow with constant flux from
smaller to larger scales, it presented a vision of turbulent flows
which was vastly different to that of the disorganized flow often
associated with three-dimensional (3D) Kolmogorov turbulence. The
inverse energy cascade allows interpretation of phenomena in
geophysical and astrophysical flows that is at odds with the picture
of turbulence of Richardson and Kolmogorov. However, as Montgomery 
and Kraichnan wrote in the concluding remarks of their famous review
\cite{Kraichnan80}, 
{\it``great caution must be used when interpreting phenomena of the
  real world in terms of asymptotic solutions of approximate
  statistical treatments of idealised theory.''} But Kraichnan and
Montgomery went beyond this warning, also suggesting that the
idealized 2D system may find its largest relevance in providing a
language for discussion of common phenomena observed in 
geophysics.

Indeed, the language and tools developed in the study of inverse
cascades have found applications in a large variety of systems. 
The occurrence of inverse cascades can be explained using statistical
mechanics in inviscid truncated systems
\cite{Kraichnan67,Kraichnan80}: when the system has two or more
quadratic conserved quantities, the solutions are not just a thermal
equilibrium between all modes, which leads to the accumulation of the
conserved quantities at small scales. Instead, other solutions can
develop involving the accumulation of one of the conserved quantities
at large scales. Moreover, this behavior is preserved in forced and
dissipative cases. To alleviate the concerns of the authors of 
Ref.~\cite{Kraichnan80}, the increase in computing power and the
improvement in experimental methods and {\it in situ} measurements 
allowed researchers to confirm these predictions, and to consider
flows in geometries or in the presence of external forces that
permitted the study of turbulence in setups that are closer to the
real world. The predictions for the two-dimensional hydrodynamic case
have been verified in experiments and in high-resolution numerical
simulations 
\cite{Clercx00,Bracco00,Kellay02,Biferale11,Boffetta12,Mininni13}. 
Inverse cascades are by now known to also take place in conducting
fluids and plasmas
\cite{Pouquet78,Ting86,Christensson01,Mininni05,Alexakis06,Mininni07},
with important consequences in space physics and astrophysics 
\cite{Demoulin09}. In atmospheric sciences, the inverse cascade plays
a  fundamental role in the study of predictability 
\cite{Lorenz69,Leith71,Leith72,Boffetta01}. And also in atmospheric
sciences, an inverse cascade of energy is known to take place in the
quasi-geostrophic (QG) equations
\cite{Charney71,Herring88,Boffetta02,Fox09}, which describe the
large-scale dynamics of atmospheric and oceanic flows. In this case,
the joint conservation of energy and of potential enstrophy is
responsible for the inverse cascade which has been also verified
numerically \cite{Vallgren10}.

The atmosphere is a rotating and stratified flow with very large
aspect ratio. While typical horizontal scales can be of the order of a 
thousand kilometers, in the vertical direction the typical height of
the troposphere is $\approx 10$ km. These features result in a flow
that can be approximated as a 2D flow at very large scales, but which
requires considering 3D rotating and stratified flows to describe in
detail small scale turbulent transport processes and non-linear
phenomena.  Compared with homogeneous and isotropic turbulence (HIT),
buoyancy forces associated with density gradients and the inertial
Coriolis force associated with the rotation of the Earth provide the
necessary restitutive forces to allow excitation of dispersive
waves. Thus, geophysical flows are often in a highly turbulent state
comprised of non-linearly interacting eddies and waves. These motions
can be classified into two classes: on the one hand, 3D modes
consisting of inertia-gravity waves evolving on a fast time scale, and
on the other hand, large-scale QG modes which evolve in a slow time
scale, and for which the Coriolis force and horizontal pressure
gradients are close to balance. \ADDB{This is the case when gravity
  and the rotation vector are  aligned, as it is the case to a good
  extent, e.g., in thin atmospheres such as the Earth's under the 
  $f$-plane approximation. In this work we will thus consider that
  gravity and the rotation vector point in the same direction. Other
  geophysical flows, such as deep atmospheres or planetary cores,
  require considering gravity and rotation pointing in different
  directions.}

The influence of rotation and stratification in the dynamics of the
atmosphere and the oceans also varies depending on the scale
studied. At the largest geophysical scales, both rotation and
stratification are significant, and the QG regime is expected to be
dominant. In the range ${\cal O}$(800-2500) km it has been observed
that the atmospheric energy spectrum scales as $E(k) \sim k^{-3}$
\cite{Muller07}, a power law consistent with the classical QG
theory of Charney \cite{Charney71}. As smaller scales are considered,
the influence of these restitutive forces on the system dynamics
decreases. Following the classical view, as the scale of interest is
decreased, the importance of rotation decreases faster than that of
stratification. At atmospheric mesoscales (horizontal scales of 
${\cal O}$(1-100) km), and in the submesoscale ocean 
(${\cal O}$(10) m to ${\cal O}$(10) km), motions are characterized by
a strong stratification with moderate rotation (with Rossby number 
$\textrm{Ro} \approx 1$, see \cite{Davidson}). At these scales, the
energy spectrum scales approximately as $E(k) \sim k^{-5/3}$
\cite{Nastrom84,Nastrom85,Muller07}. While in the atmosphere the
origin of this scaling is still unclear \cite{Sukoriansky07}, in the
ocean some evidence of an inverse cascade of energy has been found
(see, e.g., \cite{Scott05} for a study of an inverse energy cascade
from observations in the South Pacific, and \cite{Schlosser07} for
numerical simulations of an inverse energy cascade in the North
Atlantic). In the atmosphere it was suggested that this scaling
can be the result of a 2D inverse cascade fed by convective
instabilities \cite{Herring88,Verma11}. The possible coexistence
(without significant distorsions) of a direct cascade range with $E(k)
\sim k^{-3}$ fed by large scale instabilities, and of an inverse
cascade range with $E(k) \sim k^{-5/3}$ fed by instabilities at small
scales, was predicted before in \cite{Lilly83,Salmon}. However, a
$E(k) \sim k^{-5/3}$ scaling can also be observed in the direct
cascade range of rotating and stratified turbulence, and recent
atmospheric observations also seem to point to a direct cascade
process \cite{Lindborg05,Riley08}.

In fact, there is growing evidence from numerical simulations 
that there is a large variety of turbulent regimes depending on
whether geostrophic balance is broken or not, on whether QG modes
dominate over wave modes or not, and on how energy is introduced in
the system 
\cite{Smith02,Laval03,Waite04,Waite06,Sen12,Rorai13,Rorai14}. A recent
re-analysis of oceanic data, and theoretical developments in wave
turbulence theory, also indicate that there is a variety of regimes
depending on whether waves or eddies dominate
\cite{Polzin11}. Finally, the interactions between eddies, winds and
waves in the atmosphere and the oceans have dynamical consequences for
the transport and mixing of momentum, CO$_2$, and heat
\cite{Ivey08,Clark15}. Thus, it is very important to understand the
interactions between all modes in the system, and to properly
characterize the different regimes that exist in parameter space.

In the particular case of the inverse cascade range, and when
considering both rotation and stratification, several phenomena can
compete resulting in different regimes. \ADDB{While in many
  simulations with large scale forcing inverse cascades were not
  observed 
\cite{Lindborg05,Brethouwer07,Lindborg07,Aluie11,Waite11,Almalkie12,Kimura12},
  inverse cascades were reported to happen in the presence of large
  scale forcing when the forcing can excite an instability, as is the
  case of tidal forces exciting an elliptical instability studied in
  \cite{Barker13,LeReun17}.} 
With small scale forcing conclusions are somewhat contradictory, and
differ depending on whether weak rotation, weak stratification, or
rotation and stratification of comparable strength are
considered. When rotation and stratification are of comparable
strength, inverse cascades were reported
\cite{Bartello95,Metais96,Kurien08}, and were associated with the
dynamics of QG modes in the system. In purely rotating systems energy
can cascade both to the large and to the small scales
\cite{Smith96,Mininni10}, with the inverse cascade being associated
with the 2D modes of the system. For this case different scaling laws
in the inverse cascade range were also reported depending on how the
flow is stirred \cite{Sen12}. Finally, for negligible or no rotation,
although an increase of energy and of the integral vertical length
scale was reported in simulations \cite{Smith02,Laval03}, Waite and
Bartello \cite{Waite04,Waite06} conclude against the presence of an 
inverse cascade using an argument based on statistical mechanics,
similar to the one used by Kraichnan to predict the inverse cascade in
2D flows. More recently \cite{Marino14} showed, using a detailed
analysis of anisotropic fluxes, that the growth of vertical length
scales in this system is the result of highly anisotropic energy
transfers associated with the formation of  
{\it vertically sheared horizontal winds} (VSHW, see \cite{Smith02}).

The transition between these inverse cascade regimes was believed to
vary monotonically with \ADDA{the inverse Prandtl ratio $N/f$, a
  ratio measuring the strength of stratification to rotation, where
  $N$ is the Brunt-V\"ais\"ala frequency and $f$ is the Coriolis
  frequency (this parameter, and its inverse, the Prandtl ratio $f/N$,
  is named here following \cite{Vallis,Dritschel15}, and it should not
  be confused with the Prandtl number $\textrm{Pr}=\nu/\kappa$ which
  measures the ratio of kinematic viscosity to thermal diffusivity).}
Indeed, theoretical arguments suggest that the ratio of energy in
horizontal, vertical, and QG modes is governed by this ratio
\cite{Hanazaki02} (see also \cite{Bartello95,Metais96} for numerical
studies), \ADDA{and this ratio is also known to affect the large-scale
  balance in geophysical turbulence, with balance prevailing for 
  $N/f \gtrsim 1$ \cite{Dritschel15}}. Several studies showed that for
fixed rotation, increasing the stratification slows down the inverse
cascade (at least for $N/f \ge 1$). However, in a recent study
\cite{Marino13} it was shown that the inverse cascade growth speed
\ADDB{(obtained either from the time derivative of the total energy,
  or of the energy at the smallest wavenumber $k=1$)} is non-monotonic
in $N/f$, with a behavior for $N/f < 2$ different from the one found
for $N/f \ge 2$ for which the result of monotonic decrease of the
inverse cascade rate with increasing $N/f$ is recovered. In this study
it was also found that a moderate level of stratification (in the
sense that $1/2 \le N/f\le 2$) produces a faster growth of energy at
large scale than in the purely rotating case. Although linear theories
predicted a different behavior \cite{Hanazaki02}, the non-monotonicity
on $N/f$ could be expected from the theory of non-linear resonant
interactions in wave turbulence  
\cite{Cambon89,Waleffe92,Waleffe93,Cambon97,Cambon01,Waite06}.
\ADDA{It is important to note here that although atmospheric flows
  typically have an inverse Prandtl ratio $N/f$ of $10^2$ (or
  equivalently, a Prandtl ratio $f/N$ of $10^{-2}$, see
  \cite{Vallis,Dritschel15}), some flows in the ocean at mid latitude
  can have $N/f$ of order unity \cite{Nikurashin12}.}

The relevance (or not) of the wave modes in these flows can be
understood from results in the theory of resonant waves. With
sufficient rotation and stratification, non-linear interactions
between triads of wave modes are expected to become the 
predominant mechanism of energy transfer. Resonant wave theory 
predicts that given three modes with wave vectors ${\bf k}$, 
${\bf p}$, and ${\bf q}$, these can interact and transfer energy
between themselves if the wave vectors form a triangle with 
${\bf k} = {\bf p} + {\bf q}$, and if the modes are also resonant,
i.e., if their frequencies satisfy  
$\omega({\bf k}) = \omega({\bf p}) + \omega({\bf q})$
\cite{Cambon89,Waleffe93}. While this theory explains the development
of anisotropy and the tendency towards two-dimensionalization of some
flows (necessary for the development of inverse cascades), it fails to
explain how energy reaches the slow modes which seem to be the
dominant modes in the inverse cascade dynamics
\cite{Waleffe93}. Interestingly, the relevance of these interactions is
non-monotonic with increasing rotation and stratification. In the case
of rotating and stratified flows, there is a range of parameters 
$1/2 \le N / f \le 2$ for which all resonances are cancelled
out. Outside this region, the resonant triads that arise with
$N/f <1/2$ tend to two-dimensionalize the flow, forming vertical
structures in the shape of columns, while the resonant triads
arising for $N/f> 2$ tend to unidimensionalize the flow, forming
structures in the shape of pancakes \ADDB{(although a stratified flow
  could be visualized as a superposition of 2D layers, dynamically the
  strongest gradients develop in the vertical direction with very
  weak horizontal gradients)}. Thus inverse cascades can be expected
to be stronger in the former case. In between, when resonant 
interactions are not present and QG modes can be expected to be
dominant, Charney \cite{Charney71} argues that turbulence should be
isotropic in the rescaled vertical coordinate $(N/f) z$, implying
that the quotient between horizontal and vertical scales should grow
linearly with $N/f$. The non-monotonic behavior of resonant triads
with $N/f$ can therefore be expected to play a role in the development
of inverse cascades.

In this paper we first review previous results on the strength of the
inverse cascade in rotating and stratified flows, and then present new
results on the effect of varying the strength of rotation and
stratification (measured by the ratio $N/f$) on the amplitude of the
waves and on the QG behavior of the flow. We use the spatio-temporal
spectrum, and characterization of the flow temporal and spatial
scales, to show that in the range $1/2 \le N/f \le 2$ QG modes
dominate the dynamics, while the strength of the waves is weaker. The
structure of the remaining of the paper is as follows. Section
\ref{sec:Boussinesq} introduces the Boussinesq approximation, 
discusses the role of linear solutions of the equations and of the
resonant triads, and shows that resonant triads do not exist for 
$1/2 \le N/f \le 2$. Then, in Sec.~\ref{sec:inverse} we review
previous results on inverse cascades in rotating and stratified
flows. We consider the purely rotating case, the purely stably
stratified case, and parametric studies of rotating and stratified
flows as a function of $N/f$. Sections \ref{sec:qgnumerics} and
\ref{sec:spacetime} then present new results. In
Sec.~\ref{sec:qgnumerics} we present a parametric study using several
simulations at moderate spatial resolution, and show that simulations
in the range $1/2 \le N/f \le 2$ are compatible with some predictions
from QG theory. Then, in Sec.~\ref{sec:spacetime} we use the
spatio-temporal spectrum to quantify the strength of the waves and of
QG modes, and explicitly show that waves are less relevant in the same
range of $N/f$. Finally, in Sec.~\ref{sec:conclusions} we present the
conclusions.

\section{\label{sec:Boussinesq}The Boussinesq approximation}

\subsection{\label{sec:equations}The equations}

The dynamics of a stably stratified incompressible fluid subjected to
background rotation can be described, under the Boussinesq
approximation, by the momentum equation for the velocity ${\bf u}$,
and an equation for the temperature fluctuations $\theta$,
\begin{equation}
\partial_t {\bf u} + \mbox{\boldmath $\omega$} \times
    {\bf u} + 2 \mbox{\boldmath $\Omega$} \times {\bf u}  =
    - \nabla {\cal P} - N \theta \hat{z} + {\bf F} + 
    \nu \nabla^2 {\bf u} ,
\label{eq:momentum}
\end{equation}
\begin{equation}
\frac{\partial \theta}{\partial t} + {\mathbf u} \cdot \nabla \theta =
    N {\mathbf u} \cdot \hat{z} + \kappa \nabla^2 \theta ,
\label{eq:temp}
\end{equation}
together with the incompressibility condition,
\begin{equation}
\nabla \cdot {\bf u} = 0.
\label{eq:incompressible}
\end{equation}
Here gravity points in the $\hat{z}$ direction, and for simplicity we
will also consider the rotation axis in the same direction, so that
$\mbox{\boldmath $\Omega$} = \Omega \hat{z}$ where $\Omega$ is the
rotation frequency and $f=2\Omega$ the Coriolis frequency. In
Eq.~(\ref{eq:momentum}) $N$ is the Brunt-V\"ais\"al\"a frequency, 
\ADDB{$\mbox{\boldmath $\omega$} = \nabla \times {\bf u}$ is the
  vorticity,} ${\bf F}$ is an external mechanical forcing, ${\cal P}$
is the total pressure per unit of mass (including the centrifugal
acceleration and the background hydrostatic pressure), and $\nu$ is
the kinematic viscosity. In Eq.~(\ref{eq:temp}) $\kappa$ is the
thermal diffusivity (in the following equal to the kinematic
viscosity, $\kappa = \nu$, so the Prandtl number is $\textrm{Pr}=1$).

Three dimensionless numbers play an important role to characterize the
different regimes in the system. These are the Reynolds, Rossby, and
Froude numbers
\begin{equation}
\textrm{Re} = \frac{UL}{\nu} , \,\,\, 
\textrm{Ro} = \frac{U}{fL} , \,\,\, 
\textrm{Fr} = \frac{U}{NL} \,
\end{equation}
where $U$ is the r.m.s.~velocity and $L$ the flow integral
scale \ADDB{(as in flows with inverse cascades $U$ will be
  quasi-stationary at best, unless explicitly noted we will
  consider $U$ as the steady-state r.m.s.~velocity at the forced
  scale)}. While the Reynolds number measures the ratio of inertial to
viscous accelerations in the flow, the Rossby and Froude numbers are
respectively inverse measures of the relevance of rotation and of 
stratification. The \ADDA{inverse Prandtl ratio $N/f$} can be written
in terms of these numbers as $N/f=\textrm{Ro}/\textrm{Fr}$
\ADDA{\cite{Vallis,Dritschel15}}. Other dimensionless numbers are 
known to play important roles (e.g., the buoyancy Reynolds number 
$R_b = \textrm{Re}\, \textrm{Fr}^2$ 
\cite{Shih05,Brethouwer07,Ivey08,Waite11}), as well as the wave
numbers at which rotation \cite{Mininni12,Delache14} or stratification
\cite{Almalkie12,Rorai15} become negligible. \ADDA{The former is the
  Zeman wavenumber
\begin{equation}
k_\Omega = \sqrt{\frac{f^3}{\epsilon}} ,
\label{eq:Zeman}
\end{equation}
at which the period of inertial waves is the same as the eddy turnover
time, and thus isotropy is expected to be recovered in a purely rotating
flow. The latter is the Ozmidov wavenumber
\begin{equation}
k_\textrm{Oz} = \sqrt{\frac{N^3}{\epsilon}} ,
\label{eq:Ozmidov}
\end{equation}
at which the period of gravity waves is the same as the eddy turnover 
time, and thus isotropy is expected to be recovered in a purely
stratified flow. In both cases, $\epsilon$ is the energy injection
rate (which in our case is equal to the kinetic energy injection rate
$\epsilon_V = \left< {\bf u}\cdot {\bf F} \right>$, as we only force
the momentum equation).}

\begin{figure}
\includegraphics[width=8cm]{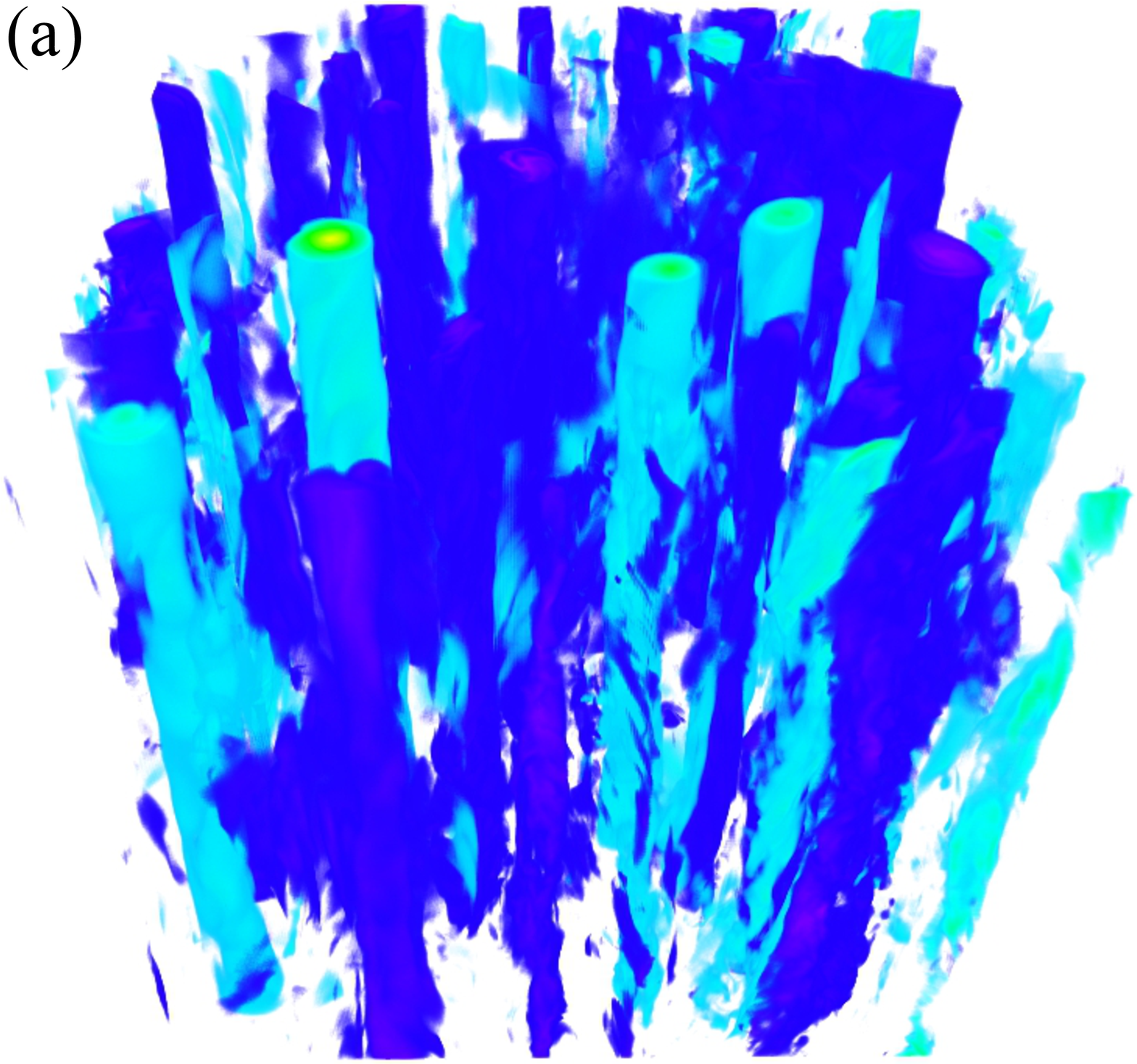}
\includegraphics[width=8cm]{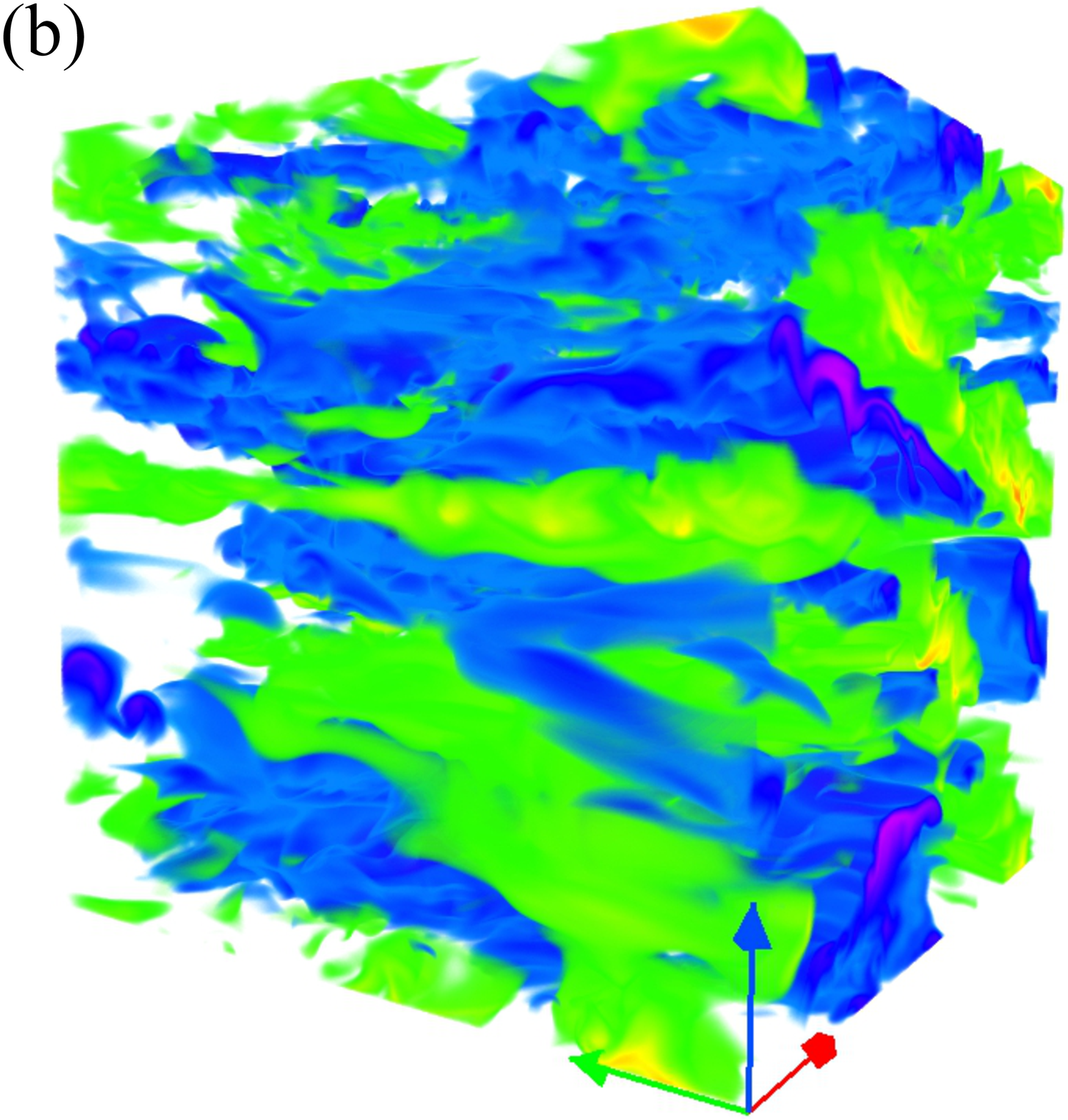}
\caption{({\it Color online}) 
  (a) Vertical velocity in a $1536^3$ simulation of rotating
  turbulence \cite{Mininni10}. Note the presence of column-like
  structures. 
  (b) Temperature fluctuations in a $1024^3$ simulation of
  rotating and stratified turbulence with \ADDB{$N/f=2$} 
  \cite{Marino13}. Note the formation of slanted layers, with an angle
  \ADDB{related} to the ratio $N/f$. In the purely stratified case,
  the layers are horizontal.} 
\label{fig:rendering}
\end{figure}

When the Reynolds number is large enough the system is in a turbulent
state. But even at very large Reynolds numbers, for $\textrm{Fr}$ and
$\textrm{Ro}$ small enough, waves are present that affect the
turbulent scaling and transport. Thus this system of equations has
been extensively studied in numerical simulations as a way to gain a
better understanding of atmospheric turbulence in a simplified set
up. As mentioned in the introduction, in the atmospheric mesoscales
(where the stratification dominates above the rotation, but the
dominant regime is quasi-geostrophic and not dominated by
inertia-gravity waves), a spectrum of energy compatible with the power
law $E(k) \sim k^{-3}$ has been reported. However, a spectrum 
$E (k) \sim k^{-5/3}$ has also been observed (see, e.g.,
\cite{Cho99}), and numerical simulations of
Eqs.~(\ref{eq:momentum})-(\ref{eq:incompressible}) have generated
results consistent with this power law in the presence of
instabilities and small-scale overturning
\cite{Vincent79,Lindborg05}. Moreover, when overturning is suppressed 
by dissipation, the spectrum in the simulations becomes steeper. The
asymptotic dynamics of this system in the limit of strongly rotating
flows generates column-like structures \cite{Cambon89}, while in the
limit of strongly  stratified vortical motions it seems to consist of
quasi-horizontal fluid layers partially decoupled between themselves
\cite{Liechtenstein05} (see Fig.~\ref{fig:rendering}). While in the
rotating case the direct cascade spectrum of 
Eqs.~(\ref{eq:momentum})-(\ref{eq:incompressible}) seems to scale as
$E(k) \sim k^{-2}$ \cite{Cambon89,Cambon01,Muller07,Mininni10},
Billant and Chomaz \cite{Billant01} argue that in the purely
stratified case the vertical characteristic scale $L_z$ scales as
$U/N$, thus suggesting that the vertical energy spectrum should follow
a power law $E(k_z) \sim N^2 k_z^{-3}$. This behavior was confirmed
in numerical simulations \cite{Lindborg05,Rorai14,Rorai15}, where it
was also observed that in the purely stratified case the parallel
spectrum $E(k_z)$ is flat for wave numbers smaller than 
$k_z \sim N/U$ \cite{Smith96,Waite04,Rorai15}. As the parallel
spectrum dominates over the perpendicular when stratification is
sufficiently strong, the isotropic spectrum also follows a law $E(k)
\sim k^{-3}$ in the purely stratified case. In the following sections
we introduce a decomposition of the flow into slow and fast modes, and
the argument of resonant triads, that allow interpretation of some of
these results in the direct cascade range, to later consider the case
of the inverse cascade.

\subsection{\label{sec:linear}Linear modes}

In the inviscid and diffusion free limit and in the absence of
external forces ($\nu = \kappa = {\bf f} = 0$), the Boussinesq
equations conserve the potential vorticity
\begin{equation}
\frac{D}{Dt} \left( \mbox{\boldmath $\omega$}_a \cdot \nabla \theta
  \right) = 0,
\end{equation}
where $\mbox{\boldmath $\omega$}_a$ is the absolute (total) vorticity
in the laboratory frame, 
$\mbox{\boldmath $\omega$}_a = \mbox{\boldmath $\omega$} + 2
\mbox{\boldmath $\Omega$}$. When rotation and the
Brunt-V\"ais\"ala frequencies are homogeneous, except for a
multiplying prefactor and a constant term, the potential vorticity can
be written as
\begin{equation}
P.V. = f \frac{\partial \theta}{\partial z} - N \omega_z +
  \mbox{\boldmath $\omega$} \cdot \nabla \theta .
\end{equation}
The conservation of potential vorticity imposes strong constraints on
the flow. As a result, solutions to the Boussinesq equations
are often decomposed into two groups of modes: inertia-gravity waves
with non-zero frequency and zero potential vorticity (also called
``fast'' modes), and modes with zero frequency and non-zero potential
vorticity (also called ``slow'' modes). By replacing
solutions
\begin{equation}
({\bf u} , \theta) = \mbox{\boldmath $\psi$} ({\bf k}) e^{i({\bf k} \cdot
  {\bf x} + \omega t)} ,
\end{equation}
in the linearized Boussinesq equations, three modes per wave vector
${\bf k}$ can be identified. The two modes with non-zero frequency
have dispersion relation \cite{Davidson}
\begin{equation}
\omega({\bf k}) = \pm \frac{\sqrt{f^2 k_\parallel^2 + N^2
  k_\perp^2}}{k},
\label{eq:disprel}
\end{equation}
where the directions parallel and perpendicular are taken with respect
to the direction of gravity and of the rotation axis, i.e.,
$k_\parallel=k_z$, $k_\perp=(k_x^2+k_y^2)^{1/2}$, and 
$k=|{\bf k}|$. The group velocity for these modes is
\begin{equation}
{\bf c} = \frac{f^2-N^2}{\omega k^4} [{\bf k} \times (k_\parallel
  \hat{z} \times {\bf k})] , 
\end{equation}
and for $f=N$ it can be expected that the role of the waves in the
flow dynamics should be less relevant. Also, note that for $f=0$ (the
purely stratified case) Eq.~(\ref{eq:disprel}) reduces to the
dispersion relation of internal gravity waves (with the modes with
$k_\perp = 0$ corresponding to the VSHW \cite{Smith02}, which have
zero frequency and are thus ``slow''). And in the case with $N=0$
(purely rotating flow), the dispersion relation reduces to that of
inertial waves (with the slow modes corresponding to 2D modes with
$k_\parallel = 0$).

In the non-linear case, the fields in Fourier space can be expanded in
terms of these three modes
\cite{Waleffe93,Babin97,Julien98,Smith99,Bellet06,Davidson,Kafiabad16},
\begin{eqnarray}
({\bf u}_{\bf k}, \theta _{\bf k}) &=& a^+ ({\bf k},t) \mbox{\boldmath
  $\psi$}^+({\bf k}) e^{i \omega t} + a^- ({\bf k},t) \mbox{\boldmath
  $\psi$}^-({\bf k}) e^{-i \omega t} \nonumber \\
{} &+& a^0 ({\bf k},t) \mbox{\boldmath $\psi$}^0({\bf k}) ,
\label{eq:decompose}
\end{eqnarray}
where $a^{(\alpha)}({\bf k},t)$ (with $\alpha = +$, $-$, $0$ labeling
the three modes) are slowly evolving amplitudes.

\subsection{\label{sec:resonant}Resonant triads}

Fourier transforming the Boussinesq equation, and using the
decomposition in Eq.~({\ref{eq:decompose}), the following equation is 
obtained \cite{Waleffe93,Babin97,Julien98}
\begin{eqnarray}
\frac{d a_{\bf k}^{(\alpha)}}{d t} &=& \sum_{{\bf k} = {\bf p}+{\bf q}}
  C_{\bf k p q}^{\alpha \beta \gamma} a_{\bf p}^{(\beta)} a_{\bf
  q}^{(\gamma)} e^{i\left (\omega_{\bf p}^{(\beta)} + \omega_{\bf
      q}^{(\gamma)} - \omega_{\bf k}^{(\alpha)} \right)} \nonumber
  \\
{} &+&  F_{\bf k}^{(\alpha)} - D_{\bf k}^{(\alpha)} ,
\label{eq:transform}
\end{eqnarray}
where $C_{\bf k p q}^{\alpha \beta \gamma}$ is a coupling coefficient
between modes, $F_{\bf k}^{(\alpha)}$ is the energy injection at
wavenumber ${\bf k}$, and $D_{\bf k}^{(\alpha)}$ is the
dissipation. The first term on the r.h.s.~of this equation corresponds
to the non-linear triadic interactions. As nonlinearities in the
Boussinesq equations are quadratic, modes are coupled with triads that
can exchange energy between themselves while conserving the total
energy. The convolution theorem (when the quadratic terms are Fourier
transformed) imposes the well known triadic condition
\begin{equation}
{\bf k} = {\bf p}+{\bf q} .
\label{eq:triad}
\end{equation}
But the presence of waves imposes an extra condition. Integrating
Eq.~(\ref{eq:transform}) in one period of the waves, the first term
    on the r.h.s.~of the equation vanishes except when
\begin{equation}
\omega_{\bf k}^{(\alpha)} = \omega_{\bf p}^{(\beta)} + \omega_{\bf
    q}^{(\gamma)},
\label{eq:resonance}
\end{equation}
which is the so-called resonant triad condition, and selects only the
triads with constructive interference between the non-linearly
interacting waves.

Waleffe describes the triadic interactions exhaustively for
homogeneous turbulence in \cite{Waleffe92}, and the resonant triadic
interactions for rapidly rotating flows in \cite{Waleffe93}, where he
presents an argument for the two-dimensionalization of rotating
flows. His ``instability assumption'' states that the statistical
direction of the energy transfer is determined by the stability of the
corresponding elemental triadic interaction. In practice, this results
in a preferential transfer of energy towards the modes with zero
frequency, although wave turbulence theories cannot explain how
energy ultimately reaches these modes \cite{Alexakis15,Clark16} 
\ADDB{(as shown in \cite{Smith05}, near-resonances are needed
  to explain the efficient transfer of energy to these modes, while
  non-resonant interactions act reducing this net energy
  transfer)}. In spite of these limitations, the argument is succesful
in explaining how anisotropy develops. As mentioned in
Sec.~\ref{sec:linear}, for purely rotating flows
Eq.~(\ref{eq:disprel}) reduces to 
$\omega({\bf k}) = \pm fk_\parallel/k$, which is the dispersion 
relation of inertial waves. Energy is then transferred preferentially
towards modes with $k_\parallel \approx 0$, which correspond to the 2D
modes, and if the energy can reach those modes it can be expected to
suffer an inverse cascade towards large scales. In purely stratified
flows Eq.~(\ref{eq:disprel}) reduces to 
$\omega({\bf k}) = \pm N k_\perp/k$ (internal gravity waves), and
energy then is transferred preferentially towards modes with 
$k_\perp \approx 0$, i.e., to modes with vertical shear. In the former
case this explains the formation of structures with small vertical
gradients (columns, see Fig.~\ref{fig:rendering}), while in the latter
case it explains the formation of pancake-like structures (see also
Fig.~\ref{fig:rendering}).

\subsection{\label{sec:char}Characteristic time scales}

In wave turbulence the presence of multiple time scales does not
allow for direct estimation of scaling laws in the inertial range as
is often done in Kolmogorov theory of turbulence. While in HIT there
is only one time scale in the inertial range (the eddy turnover time),
in the presence of waves this time scale coexists with the period of
the waves, precluding the construction of a unique time scale on
dimensional grounds. While in weak wave turbulence regimes 
this problem can be circumvented (see, e.g., \cite{Nazarenko}), in the
strong turbulent case the problem persists. Interestingly, one of the
first advances towards the construction of phenomenological theories
for this kind of systems was also done by Kraichnan, who put forward a
theory for the interaction of eddies and Alfv\'en waves in
magnetohydrodynamic turbulence \cite{Kraichnan65}.

Thus, it is important to identify the different time scales relevant
in our problem. The decomposition introduced in Sec.~\ref{sec:linear}
is useful to this end. The first time scale will naturally be
proportional to the wave period
\begin{equation}
\tau_w ({\bf k}) = \frac{C_w}{\omega ({\bf k})} = \frac{C_w
    k}{\sqrt{f^2 k_\parallel^2 + N^2 k_\perp^2}} ,
\label{eq:omegaig}
\end{equation}
where $C_w$ is be a dimensionless constant of ${\cal O}$(1) which can
be obtained directly from the auto-correlation function of the Fourier
modes of the velocity and temperature fields (see \cite{Clark14b}).

From Eq.~(\ref{eq:transform}) it is to be expected that the fastest
time dominates the flow dynamics and the energy transfer. However,
although for small enough $\textrm{Ro}$ and $\textrm{Fr}$ the waves
are expected to be fast, the period of the waves is not homogeneous in
Fourier space as it depends \ADDB{(only) on the direction of} 
${\bf k}$. Thus, different regions of Fourier space can be dominated
by different time scales depending on what mode is the fastest. The
period of the waves must then be compared with the eddy turnover time
\begin{equation}
\tau_{nl} (k) = \frac{C_{nl}}{k \sqrt{k E(k)}} ,
\label{eq:nltime}
\end{equation}
where $C_{nl}$ is another dimensionless constant of ${\cal O}$(1). In
principle $\tau_{nl}$ depends on the amplitude of ${\bf k}$ and on its
direction in Fourier space, i.e., $\tau_{nl} ({\bf k})$, since the
energy spectrum in these systems is anisotropic. But for simplicity we
will use here the isotropic energy spectrum $E(k)$ to estimate the
non-linear time. \ADDA{The condition 
  $\tau_{nl} (k) = \tau_w ({\bf k})$ for either $N=0$ or $f=0$ yields
  respectively the Zeman and Ozmidov wave numbers defined in
  Eqs.~(\ref{eq:Zeman}) and (\ref{eq:Ozmidov}). For larger wave
  numbers, the role of the waves in the non-linear energy transfer can
  be expected to be negligible.}

The last relevant time scale is that of the sweeping of the small
eddies by the large scale flow, which becomes the dominant
Eulerian decorrelation mechanism as soon as its characteristic time
scale becomes the fastest of the three (\ADDA{this often happens at
  wave numbers smaller than the Zeman and Ozmidov wave numbers, see
  \cite{Clark14b,Clark15}}; Kraichnan played an important role in
highlighting the relevance of this time scale in the case of isotropic
and homogeneous turbulence \cite{Chen89}). The time to sweep an eddy 
of size $\sim 1/k$ by a large scale flow with amplitude $U$ is simply
\begin{equation}
\tau_{sw} (k) = \frac{C_{sw}}{U k} ,
\label{eq:sweeping}
\end{equation}
where $C_{sw}$ is another dimensionless constant of ${\cal O}$(1) that
can be determined from the data. These time scales will be very
important in the following sections.

Using combinations of these time scales, phenomenological theories for
the direct cascade spectrum can be also constructed, following the
ideas of Kraichnan \cite{Kraichnan65}. These phenomenological
arguments yield spectra compatible with the scaling laws $\sim k^{-2}$
and $\sim k^{-3}$ discussed in the introduction and in 
Sec.~\ref{sec:equations}, for solutions of
Eqs.~(\ref{eq:momentum})-(\ref{eq:incompressible}), depending on the
strength of rotation and stratification considered. For more details 
see \cite{Dubrulle92,Zhou95,Billant01,Pouquet10}. \ADDB{These time
  scales, and the results presented in the next two subsections, will
  be important to elucidate the results of the spatio-temporal
  analysis in Sec.~\ref{sec:spacetime}.}

\subsection{\label{sec:nonresonant}Non-resonant range}

It can be shown that in the range $1/2 \le N/f \le 2$ there are no
resonant triadic interactions, and therefore all triadic interactions
must be near-resonant or non-resonant. \ADDB{In this section we
  summarize the argument given in \cite{Smith02}.} As discussed in 
Sec.~\ref{sec:resonant}, non-linear interaction of inertia-gravity
waves require that the interacting modes ${\bf k}$, ${\bf p}$, and 
${\bf q}$ form a triangle, and that the sum of their frequencies give
constructive interference. If $N=f$, it is then trivial to see from
Eq.~(\ref{eq:disprel}) that
$\omega ({\bf k}) = \pm N$ for all ${\bf k}$, and
then
\begin{equation}
\omega ({\bf p}) + \omega ({\bf q}) -\omega ({\bf k}) = \pm N, \pm
  2N, \textrm{or} \pm 3N .
\end{equation}
Therefore, the sum of the three frequencies cannot be zero and the
condition given by Eq.~(\ref{eq:resonance}) cannot be
fulfilled. \ADDB{Thus, resonant triads do not exist in this case.}

In the most general case, Eq.~(\ref{eq:disprel}) implies that
\begin{eqnarray}
\min \{|\omega ({\bf k})|\} &=& \min \{N,f\} , \\
\max \{|\omega ({\bf k})|\} &=& \max \{N,f\} .
\end{eqnarray}
Assuming without loss of generality
\ADDB{$\omega ({\bf k})$, $\omega ({\bf p})$, $\omega ({\bf q}) \ge 0$},
to satisfy Eq.~(\ref{eq:resonance}) we need 
$2 \min \{ N,f \} < \max \{ N,f \}$. When $\max \{ N,f \} = N$, we then
need $N/f > 2$. When $\max \{ N,f \} = f$, then $N/f < 1/2$. In these
cases there are triads which can satisfy the resonant
condition. However, in the range
\begin{equation}
\frac{1}{2} \le \frac{N}{f} \le 2,
\end{equation}
there is again no triad that can be resonant. \ADDB{Only
  near-resonances and non-resonant triads are then left to transfer
  energy between scales.}

It is now clear that this range separates two regimes with different
behavior. It is expected that outside this range, but in its vicinity,
very few resonant triads should be available (thus, resonant triads
should play a more relevant role for $N/f \ll 1/2$ or $N/f \gg
2$). For $N/f <1/2$, the first resonant triads that arise \ADDB{(as
  $N/f$ is decreased}) consist of a vertical mode $(0,0, k_\parallel)$
and of two quasi-horizontal modes ${\bf p}$, ${\bf q}$ with 
$p_\perp = q_\perp \gg | k_\parallel |, |p_\parallel |,  |q_\parallel|$.
Waleffe's instability assumption then suggests that energy should go
from the vertical mode to the perpendicular (slow) modes. Similarly,
for $N / f> 2$, the first resonant triads \ADDB{that appear as $N/f$
  is increased} consist of a horizontal mode and two quasi-vertical
modes. In this case, the transfer should be expected to occur from the
horizontal modes to the quasi-vertical (slow) modes. As with the
arguments in Sec.~\ref{sec:resonant}, these arguments suggest that the
resonant triads for $N/f <1/2$ tend to bi-dimensionalize the flow
\ADDB{(i.e., to make horizontal gradients much larger than vertical
  gradients, as energy is preferentially transferred towards modes
  with $k_\perp \gg k_\parallel$)}. On the contrary, the resonant
triads for $N/f> 2$ tend to unidimensionalize the flow \ADDB{(i.e.,
  to make vertical gradients much larger than horizontal gradients, or
  $k_\parallel \gg k_\perp$). An illustration of this can be seen in
  Fig.~\ref{fig:rendering}, which shows column-like structures in the
  vertical velocity in a simulation with pure rotation, and
  pancake-like structures in a simulation with rotation and
  stratification.}

\subsection{\label{sec:qg}Quasi-geostrophic equation}

Simulations in the range $1/2 \le N/f \le 2$ show that at scales
larger than the forced scales QG modes dominate the dynamics 
\cite{Smith96}, in agreement with the theoretical prediction that
resonant triads are not available in this region of parameter
space. \ADDB{As in other limits of geophysical flows,} this can be
used to derive a reduced system of equations. Although the QG
approximation is  often derived in the limit of very small Froude
number and moderate rotation (of interest in atmospheric sciences),
Smith and Waleffe \cite{Smith96,Smith02} showed that the QG equations
can be also obtained for moderate values of $N/f$, which can be of
interest for some oceanographic applications. \ADDB{Here we briefly
  summarize their derivation.}

If only modes with zero frequency are excited, in the inviscid and
unforced case Eq.~(\ref{eq:transform}) can be written as
\begin{equation}
\frac{d a_{\bf k}^0}{dt} = \sum_{{\bf k} = {\bf p} + {\bf q}} C_{{\bf
      k}{\bf p}{\bf q}} a_{\bf p}^0 a_{\bf q}^0 ,
\label{eq:stepqg}
\end{equation}
where 
\cite{Smith02} \ADDB{
\begin{eqnarray}
&& C_{{\bf k}{\bf p}{\bf q}} \equiv C_{{\bf k}{\bf p}{\bf q}}^{000} = \\
&& \frac{i N ({\bf p} \times {\bf q}) \cdot \hat{z}
    \left[f^2(q_\parallel^2-p_\parallel^2)+
          N^2(q_\perp^2-p_\perp^2)\right]} 
          {\left[(f^2k_\parallel^2+N^2k_\perp^2)
            (f^2p_\parallel^2+N^2p_\perp^2)
            (f^2q_\parallel^2+N^2q_\perp^2)\right]^{1/2}} . \nonumber
\end{eqnarray}
The triadic condition in Eq.~(\ref{eq:triad}) implies that 
$({\bf p} \times {\bf q}) \cdot \hat{z} = ({\bf q} \times {\bf k})
  \cdot \hat{z} = ({\bf k} \times {\bf q}) \cdot \hat{z}$, that is,
that the $C_{{\bf k}{\bf p}{\bf q}}$ coupling coefficients are cyclic
permutations of the wave vectors multiplied by 
$f^2(q_\parallel^2-p_\parallel^2)+N^2(q_\perp^2-p_\perp^2)$. It then
follows that
\begin{equation}
C_{{\bf k}{\bf p}{\bf q}} + C_{{\bf p}{\bf q}{\bf k}} + C_{{\bf q}{\bf
    p}{\bf k}} = 0,
\end{equation}
and
\begin{eqnarray}
&& (f^2k_\parallel^2+N^2k_\perp^2) C_{{\bf k}{\bf p}{\bf q}} + 
  (f^2p_\parallel^2+N^2p_\perp^2) C_{{\bf p}{\bf q}{\bf k}} +
     \nonumber \\
&& (f^2q_\parallel^2+N^2q_\perp^2) C_{{\bf q}{\bf p}{\bf k}} = 0,
\end{eqnarray}
which convey two conservation laws as discussed below.}

Taking $\psi_{\bf k} = -N a_{\bf k}^0/(k \omega_{\bf k})$, and Fourier
transforming Eq.~(\ref{eq:stepqg}) into real space, we obtain
\begin{equation}
\left( \frac{\partial}{\partial t} + {\bf v} \cdot \nabla \right)
  \left( \nabla_\perp^2 + \frac{f^2}{N^2}\frac{\partial^2}{\partial
      z^2} \right) \psi({\bf x},t) = 0,
\end{equation}
where $\nabla_\perp^2 = \partial_x^2 + \partial_y^2$, 
${\bf v} = \hat {z} \times \nabla \psi$, and 
$\theta = -(f/N) \partial_z \psi$. This is the quasi-geostrophic
equation, which conserves two quadratic invariants: the total energy
\begin{equation}
E = \frac{1}{2} \left< \left|\nabla_\perp \psi \right|^2 +
  \frac{f^2}{N^2} \left(\frac{\partial \psi}{\partial z}\right)^2
  \right> ,
\end{equation}
with $\nabla_\perp = \partial_x \hat{x}+ \partial_y \hat{y}$, and the
potential enstrophy
\begin{equation}
F = \frac{1}{2} \left< \left(\nabla_\perp^2 \psi +
    \frac{f^2}{N^2}\frac{\partial^2 \psi}{\partial z^2} \right)^2
    \right> .
\end{equation}
As in the 2D case studied by Kraichnan \cite{Kraichnan67}, the
existence of two invariants in this case also allows for the
development of an inverse cascade of energy
\cite{Charney71,Herring88,Vallgren10}. \ADDB{This result will be
  important in the following sections, as we will see that in the
  range $1/2 \le N/f \le 2$, rotating and stratified turbulence can
  develop a very efficient inverse cascade of energy, with a
  prevalence of slow modes satisfying QG scaling.}

\section{\label{sec:inverse}Inverse cascades}

To discuss previous results in inverse cascades, and to analyze the
new results in this paper, we will use isotropic and anisotropic
spectra and fluxes. We thus begin by defining the isotropic kinetic
energy spectrum, which is computed in numerical simulations as
\begin{equation}
E_V(k) = \frac{1}{2} \sum_{k\le |{\bf k}|< k+1} |{\bf u}({\bf k})|^2,
\label{eq:eiso}
\end{equation}
i.e., as the energy in spherical shells of width $\Delta k = 1$ in
Fourier space. An equivalent definition is obtained for the
\ADDB{power spectrum of temperature fluctuations $E_P(k)$ (sometimes
  also called the potential energy spectrum)} replacing 
$|{\bf u}({\bf k})|^2$ by  $|\theta({\bf k})|^2$, and the total energy
spectrum is then simply constructed as $E(k) = E_V(k) + E_P(k)$. The
kinetic energy spectrum can be further decomposed into the kinetic
energy in horizontal fluctuations and in vertical fluctuations, 
$E_V(k) = E_\perp(k) + E_z(k)$, where $E_\perp(k)$ is the energy in
the $x$ and $y$ components of the velocity field, and $E_z(k)$ is the
energy in $u_z$.

As the flows we consider are anisotropic, it is useful to define the
axisymmetric kinetic energy spectrum 
\begin{equation}
e_V(k_\perp,k_\parallel)= \frac{1}{2}
    \sum_{\substack{
          k_{\perp}\le |{\bf k}\times \hat {\bf z}| < k_{\perp}+1 \\
          k_{\parallel}\le k_z < k_{\parallel}+1}} |{\bf u}({\bf
          k})|^2 ,
\label{eq:etheta}
\end{equation}
The axisymmetric kinetic energy spectrum in Eq.~(\ref{eq:etheta}) is
such that the total kinetic energy in 2D modes is 
$E_{2D} = \sum_{k_\perp} e_V(k_\perp,k_\parallel=0)$. As a result, we
will refer to $e_V(k_\perp,k_\parallel=0)$ as the kinetic energy
spectrum of the 2D modes. \ADDB{As for the isotropic spectra, we can
  decompose the axisymmetric kinetic energy spectrum into 
$e_V(k_\perp,k_\parallel)=e_\perp(k_\perp,k_\parallel)+e_z(k_\perp,k_\parallel)$,
  and we can write the axisymmetric total energy spectrum as
  $e(k_\perp,k_\parallel)=e_V(k_\perp,k_\parallel)+e_P(k_\perp,k_\parallel)$.}
Note in Eqs.~(\ref{eq:eiso}) and (\ref{eq:etheta}) we use sums as we
are considering the discrete Fourier transform for a triple
$2\pi$-periodic spatial domain. In the most general case these sums
are replaced by integrals.

To quantify energy scaling in the parallel and perpendicular
directions, reduced perpendicular and parallel spectra can then be
defined from the axisymmetric energy spectrum as, e.g., for the total
energies:
\begin{equation}
E(k_\perp) = \sum_{k_\parallel} e(k_\perp,k_\parallel) ,
\end{equation}
and
\begin{equation}
E(k_\parallel) = \sum_{k_\perp} e(k_\perp,k_\parallel) .
\end{equation}

Identification of inverse cascades requires not only the study of
the growth of energy in these spectra for small wave numbers, but also
quantification of negative flux of energy in a range of
wave numbers. To this end we can define anisotropic transfer functions
associated with each energy spectra $E(k)$, $E(k_\perp)$, and
$E(k_\parallel)$ respectively as
\begin{equation}
T(k) = - \sum_{k\le |{\bf k}|< k+1} t({\bf k}) ,
\end{equation}
\begin{equation}
T(k_\perp) = - \sum_{k_\perp \le |{\bf k}\times \hat{z}|< k_\perp+1}
  t({\bf k}) , 
\end{equation}
\begin{equation}
T(k_\parallel) = - \sum_{k_\parallel \le k_z < k_\parallel}
  t({\bf k}) ,
\end{equation}
where $t({\bf k}) = {\bf u}^\star({\bf k}) \cdot \widehat{({\bf u} \cdot
  \nabla {\bf u})} +\theta ^\star({\bf k}) \widehat{({\bf u} \cdot
  \nabla \theta)} + c.c.$, 
where $^\star$ and $c.c.$ denotes complex conjugate, and the
superscript $\widehat{}$ denotes the Fourier transform. From these
functions anisotropic fluxes are then defined as follows:
\begin{equation}
\Pi(k) = -\sum_{k'=0}^k T(k') ,
\end{equation}
\begin{equation}
\Pi(k_\perp) = -\sum_{k_\perp'=0}^{k_{\perp}} T(k'_\perp),
\end{equation}
\begin{equation}
\Pi(k_\parallel) = -\sum_{k'_\parallel=0}^{k_\parallel}
  T(k'_\parallel) .
\end{equation}
These fluxes measure energy per unit of time that goes in Fourier
space respectively across spheres with radius $k$, across cylinders
with radius $k_\perp$, and across planes with height $k_\parallel=$
constant.

\subsection{Rotating case}

In the purely rotating case with moderate Rossby number, the
preferential transfer of energy towards 2D modes results in the
development of an inverse cascade which shares some similarities with
the inverse cascade originally proposed by Kraichnan. By now evidence
of this inverse cascade has been found in numerical simulations in
finite domains \cite{Sen12} and in experiments
\cite{Campagne14}. There is still a controversy of whether this
inverse cascade can take place in infinite domains and for very small
Rossby numbers, for more details see, e.g., Ref.~\cite{Cambon04}.

An interesting property of the inverse cascade of energy in this case
is that changes in the forcing function or in the dominant time scale
can generate large differences in the scaling of the energy spectrum
(see \cite{Sen12}). In particular, the anisotropy of the forcing seems
to play an important role in setting the shape of the inverse cascade
energy spectrum. As an illustration, Fig.~\ref{fig:spectrum_rot} shows
the reduced perpendicular spectrum $E(k_\perp)$ and the horizontal
kinetic energy spectrum of 2D modes $e_\perp(k_\perp,k_\parallel=0)$
for two simulations of rotating turbulence forced at small scales.

The two simulations, which have $\textrm{Re} \approx 400$ and
$\textrm{Ro} = 0.045$, \ADDA{were continued for over 250 turnover
  times, and both have $k_\Omega\gtrsim 1800$, much larger than the
  largest resolved wave number (i.e., all wave numbers in the
  simulation are dominated by rotation).} The difference between these
two simulations is in the isotropy (or anisotropy) of the forcing
function. While in both cases a random forcing is used, in one case
energy is injected isotropically in all modes in the spherical shell
with $k=k_F \approx 40$, while in the other energy is injected mostly
in the modes (in the same spherical shell) that are close to the
$k_\perp$ plane (i.e., preferentially into modes with small
$k_\parallel$, or the 2D modes). A detailed analysis of anisotropic
fluxes in \cite{Sen12} shows that while in the former case this
results in a transfer of energy from the 3D modes to the 2D modes
which then drive the inverse cascade, in the latter case the energy
that is fed directly into the 2D modes undergoes an inverse cascade
and then feeds the 3D modes once the energy accumulates at the largest
available scale. \ADDB{Eventually, at sufficiently long times the
  energy accumulates in both cases at the smallest available wave
  number, $k=1$, and energy then leaks from these modes towards the 3D
  modes. The behavior of the systems at long times also depends on
  whether large-scale friction is used or not.}

Before this time, the changes in the energy fluxes (and in the
coupling between 2D and 3D modes) result in two different scaling
laws, with one case displaying spectra $E(k_\perp)$ and 
$ e_\perp(k_\perp,k_\parallel=0) \sim k_\perp^{-3}$, and the other
with spectra $E(k_\perp)$ and $ e_\perp(k_\perp,k_\parallel=0)$
compatible with $\sim k_\perp^{-5/3}$. These two scaling laws can be
explained using phenomenological arguments similar to those used by
Kraichnan \cite{Sen12}. In the latter case, in which energy is
injected directly into the 2D modes, and in which little energy goes
into the 2D modes from the 3D modes, we can assume that the inverse
cascade in the slow 2D modes is dominated by the turnover 
time $\tau_\perp \sim l_\perp/u_\perp$ (where $l_\perp$ is a
characteristic perpendicular length, and $u_\perp$ is the 2D
r.m.s.~velocity at that length scale). With only one relevant
timescale, Kraichnan phenomenology for the inverse cascade tells us
that the energy flux in the 2D modes goes as
\begin{equation}
\Pi_{2D} \sim \frac{u_\perp^2}{\tau_\perp} \sim
  \frac{u_\perp^3}{l_\perp} ,
\end{equation}
which results in a $\sim \epsilon^{2/3} k_\perp^{-5/3}$ spectrum for
the 2D modes (assuming the energy injection rate at all modes
$\epsilon$ and the 2D energy flux $\Pi_{2D}$ are proportional).

\begin{figure}
\includegraphics[width=8cm]{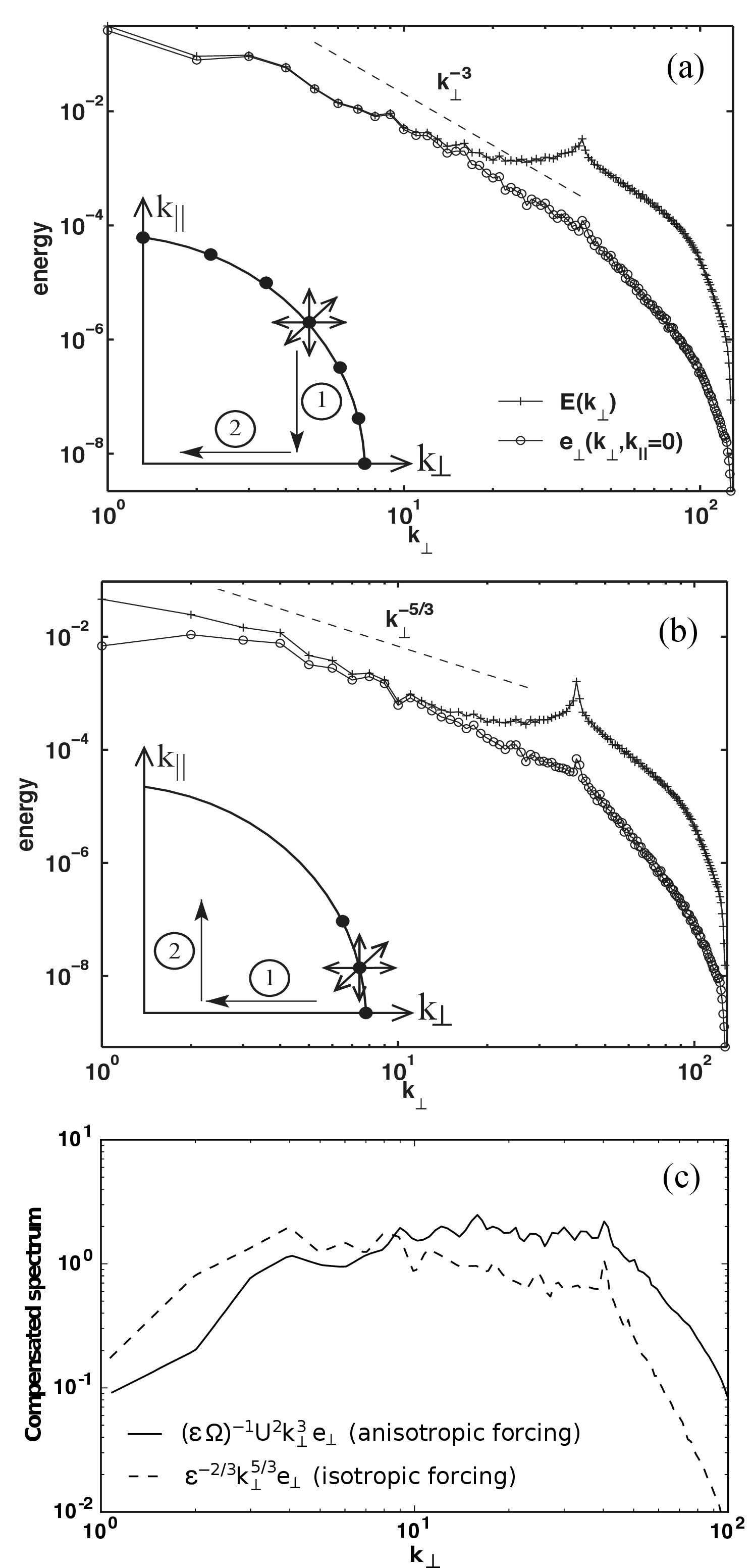}
\caption{(a) Reduced perpendicular energy spectrum $E(k_\perp)$ 
  and perpendicular kinetic energy in the 2D modes
  $e_\perp(k_\perp,k_\parallel=0)$ in a simulation of rotating
  turbulence forced at small scales with random isotropic forcing. The
  inset illustrates how the forcing acts over the spherical shell in
  Fourier space; energy is transfered first towards the 2D modes and
  then develops an inverse cascade. (b) Same for a simulation with
  anisotropic forcing. The inset illustrates how the energy that is
  injected directly into the 2D modes has an inverse cascade, and then
  it is transferred towards 3D modes as accumulation of energy at
  large scales saturates. In both figures power laws are shown as a
  reference (see \cite{Sen12} for more details). \ADDA{(c) Energy
    spectrum of 2D motions $e_\perp(k_\perp,k_\parallel=0)$
    compensated by $\epsilon^{2/3} k_\perp^{-5/3}$ in the run with
    random isotropic forcing, and by 
    $\epsilon \Omega U^{-2} k_\perp^{-3}$ (where $U$ is the
    r.m.s.~velocity at the forcing scale) in the run with anisotropic
    forcing. The amplitudes of both spectra when compensated (i.e.,
    the ``Kolmogorov constants'') are found to be of order unity.}}
\label{fig:spectrum_rot}
\end{figure}

On the other hand, if the forcing is isotropic and energy goes from
the 3D modes to the 2D modes, interactions with the 3D modes cannot be
neglected. Besides the slow turnover time $\tau_\perp$, we now have to
consider the time scale of the waves (which is the fastest time scale
for many of the 3D modes), and which we can estimate as 
$\tau_\Omega \sim 1/\Omega$. Using Kraichnan phenomenology for
interactions of waves and eddies \cite{Kraichnan65}, we can assume
that the non-linear transfer will be slowed down by a factor
proportional to the Rossby number, i.e., to the ratio of time scales
between the wave period and some non-linear turnover time. Then,
 \begin{equation}
\Pi_{2D} \sim \left( \frac{u_\perp^2}{\tau_{nl}} \right)
  \left( \frac{\tau_\Omega}{\tau_{nl}} \right)
  \sim \frac{u_\perp^2}{\Omega \tau_{nl}^2} .
\end{equation}
Assuming again $\Pi_{2D}$ is proportional to $\epsilon$, and if the
turnover time in the above expression is built upon the velocity at
the forcing scale $U$ (i.e., $\tau_{nl} \sim l_\perp/U$, assuming most
of the energy in the 2D modes comes directly from the 3D forced
modes), this relation results in a spectrum 
$\sim \epsilon \Omega U^{-2} k_\perp^{-3}$ for the 2D modes
\cite{Sen12}. \ADDA{Figure \ref{fig:spectrum_rot}(c) shows the
  spectrum of 2D modes in the simulations with isotropic and with
  anisotropic random forcing compensated by their corresponding
  expressions. In both cases, the compensated spectra have amplitudes
  close to unity in the inverse cascade range. In comparison, the
  Kolmogorov constant for the classical 2D inverse cascade is in the
  range $\approx 5.5$ -- $7$ \cite{Paret97}, although in some
  experiments its value was also reported to depend on the rate at
  which energy was extracted at large scales \cite{Sommeria86}.}

\begin{figure}
\includegraphics[width=8.5cm]{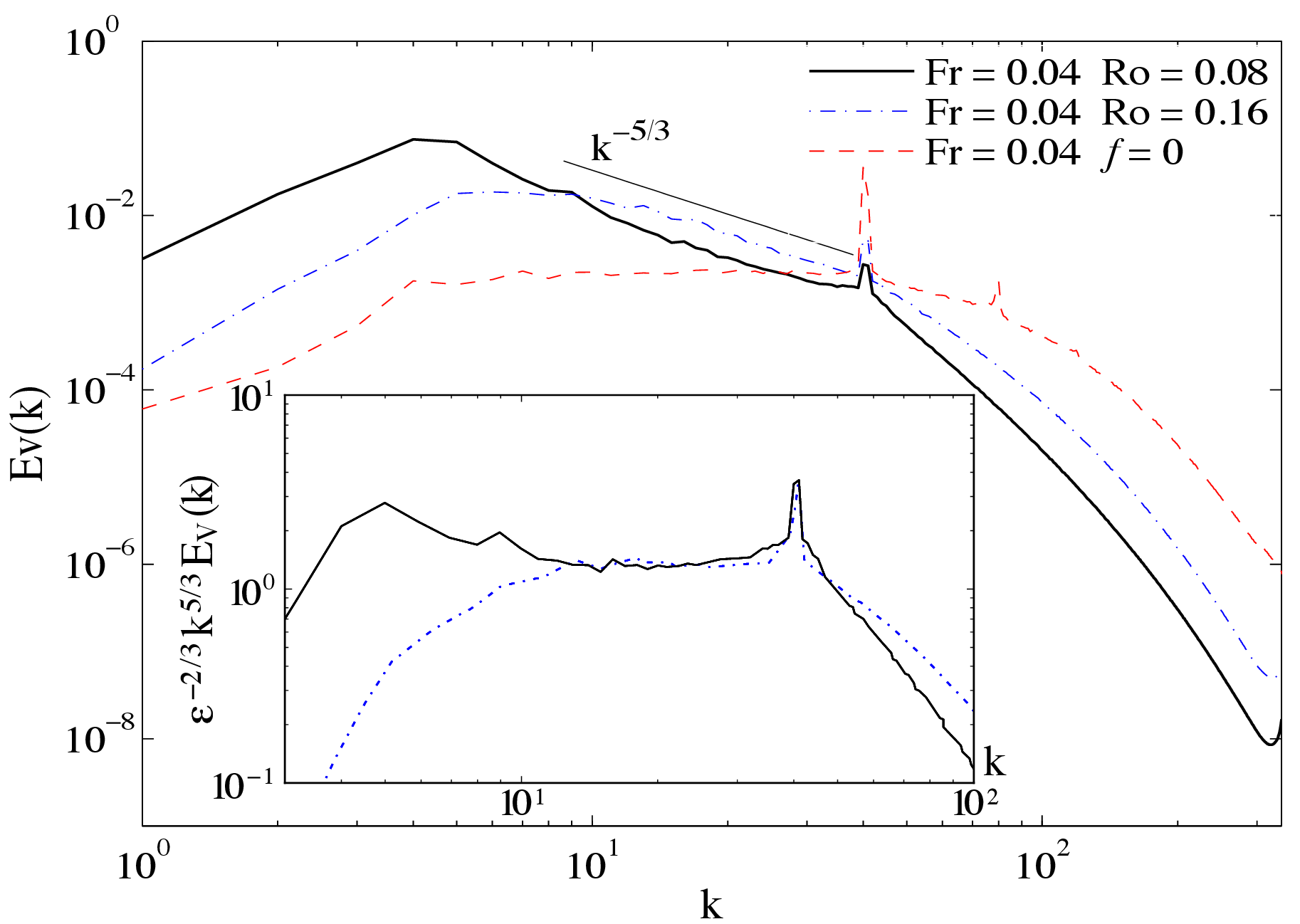}
\caption{({\it Color online})
  Isotropic kinetic energy spectrum in a $1024^3$ simulation of
  a stably stratified flow ($f=0$, i.e., without rotation), and in two
  $1024^3$ simulations of rotating and stratified flows with the same
  Froude number and different Rossby numbers (see \cite{Marino14} for
  details). Note the growth of energy at wave numbers smaller than the
  forcing wavenumber $k_F \approx 40$ in the two latter simulations,
  and the flat spectrum at the same wave numbers in the former case. A
  $\sim k^{-5/3}$ power law is indicated as a reference. \ADDA{The
    inset shows the spectrum of the two simulations with rotation and
    stratification compensated by $\epsilon^{2/3} k^{-5/3}$. The
    amplitude of the compensated spectra in the flat range (i.e., the
    ``Kolmogorov constant'') is $\approx 1.3$. Note this value is
    smaller than the Kolmogorov constant for the classical 2D inverse
    cascade $\approx 5.5$ -- $7$ \cite{Paret97}.}}
\label{fig:spectrum_rotstrat}
\end{figure}

\subsection{Stratified case}

A detailed study of the effect of the forcing on the development and
growth of large scales in stratified flows (as well as in rotating and
stratified flows) is still partially missing. For the sake of
simplicity, from here on we will consider either isotropic forcing
functions, or Taylor-Green forcing which does not inject energy
directly into 2D modes, and which for the rotating case was shown to
give an inverse cascade with the same scaling as isotropic forcing 
\cite{Sen12}.

Figure \ref{fig:spectrum_rotstrat} shows the isotropic energy spectrum
in a $1024^3$ simulation of a purely stratified flow with 
$\textrm{Re} \approx 1000$ and $\textrm{Fr} = 0.04$ ($f=0$ as there is
no rotation, \ADDA{and the Ozmidov wavenumber is 
  $k_\textrm{Oz} \approx 510$, larger than the largest resolved wave
  number in the simulation $1024/3 \approx 341$;} see
\cite{Marino13,Marino14} for more details). The flow is forced at
$k_F\approx 40$ so there is room at small wave numbers for an inverse
cascade to develop. Instead, what we observe is that energy grows at
wave numbers $k<k_F$, but with a flat spectrum in this range.

In the purely stratified case there is growth of energy at large
scales as energy piles up into VSHW, i.e., modes for the horizontal
velocity with $k_\perp \approx 0$ and with strong vertical gradients 
\cite{Smith02}. However, this growth is not accompanied by a negative
constant energy flux in a wide range of scales as expected for an
inverse cascade \cite{Waite04,Waite06,Marino14}. Instead, the
isotropic energy flux tends to be small at large scales (small
wave numbers), while the parallel energy flux becomes negative and the
perpendicular energy flux becomes positive \cite{Marino14}. In other
words, the growth of energy at large scales is not the result of a
self-similar cascade but rather of a very strong anisotropization of
the flow. Candidates for the generation of the VSHW include resonant
triads \cite{Smith02} and absorption of waves by critical layers
\cite{Clark15}.

\subsection{Rotating and stratified case\label{sec:rotstrat}}

\begin{figure}
\includegraphics[width=8.5cm]{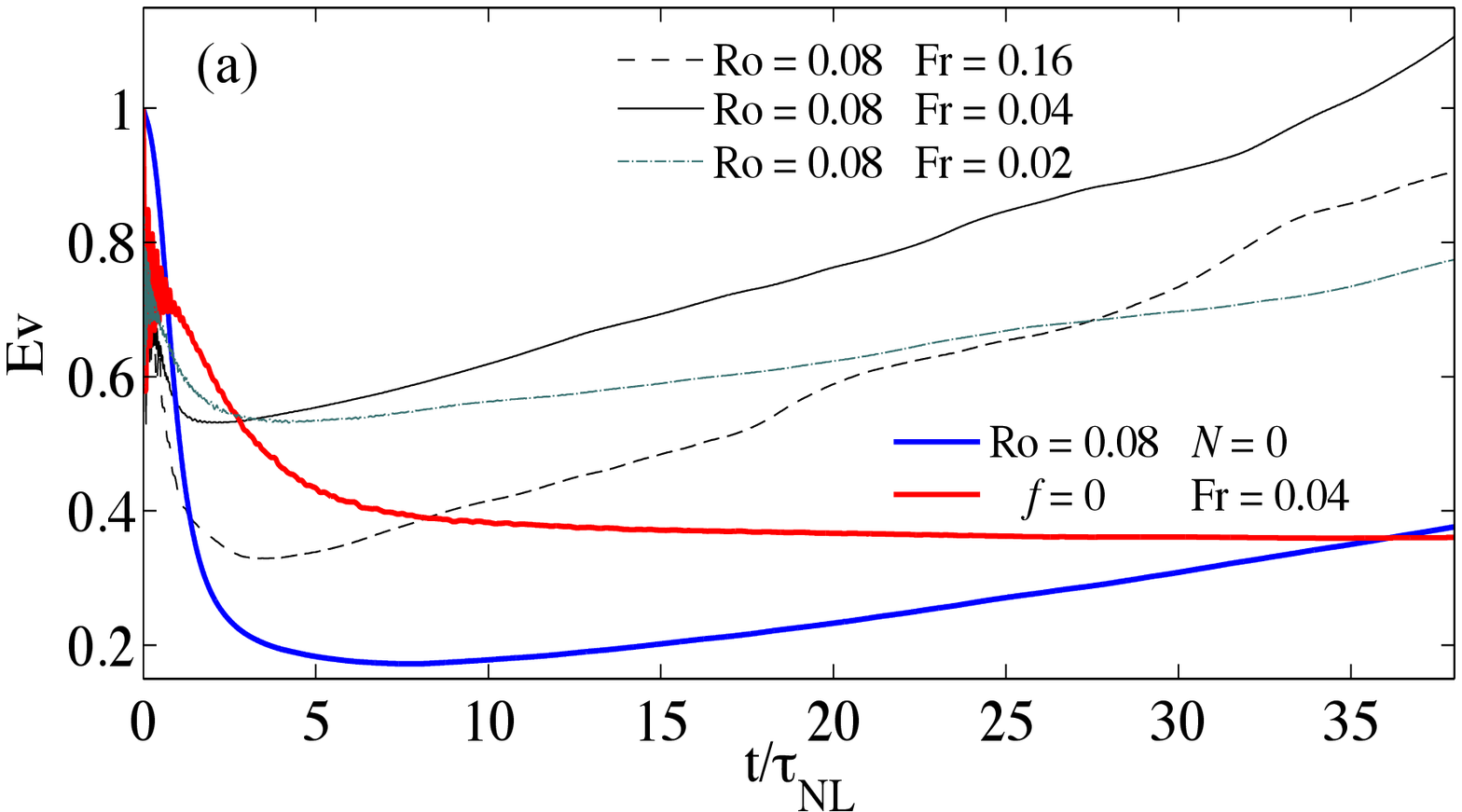}
\includegraphics[width=8.5cm]{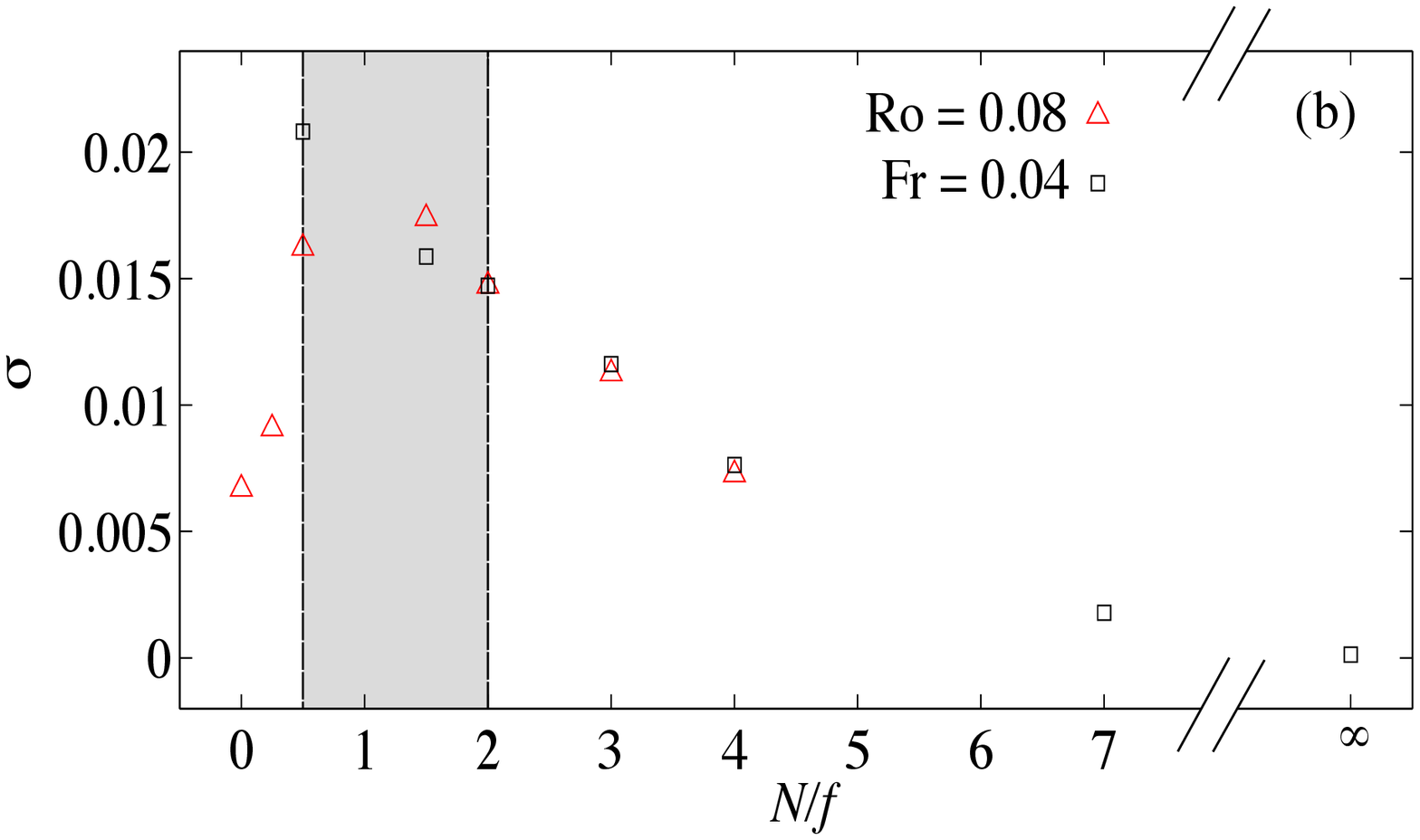}
\caption{({\it Color online}) 
  (a) Kinetic energy as a function of time in several $1024^3$
  simulations of rotating and stratified turbulence. The thick lines
  correspond to purely rotating or purely stratified simulations. The
  thin lines correspond to rotating and stratified runs. The two runs
  in the range $1/2 \le N/f \le 2$ have the fastest growth of
  $E_V(t)$. (b) Average in time growth speed $\sigma=dE_V/dt$ of the
  kinetic energy as as a function of $N/f$ for two sets of
  simulations: one with fixed $\textrm{Ro}=0.08$ and varying 
  $\textrm{Fr}$, and another with fixed $\textrm{Fr}=0.04$ and varying
  $\textrm{Ro}$. The range $1/2 \le N/f \le 2$, which displays the
  fastest growth, is indicated by the gray shading (see \cite{Marino13}
  for more details).}
\label{fig:marino}
\end{figure}

As explained in the introduction and in Sec.~\ref{sec:Boussinesq}, the
\ADDA{inverse Prandtl ratio} $N/f$ is expected to play an important
role controlling the different regimes of the inverse cascade in
rotating and stratified flows. At this point, we know that while in
the purely rotating case ($N/f \to 0$) there is an inverse cascade, in
the purely stratified case ($N/f \to \infty$) there is none. Although
{\it a priori} a monotonous decrease of the strength of the inverse
cascade could be expected between these two limit cases, the
non-monotonicity in the strength of the resonant interactions with
$N/f$ suggests there can be a distinct behavior in the range 
$1/2 \le N/f \le 2$.

This problem was considered in detail in \cite{Marino13}, where a
parametric study of the inverse cascade was done varying $N/f$ in a
large set of numerical simulations with spatial resolutions with up to
$1024^3$ grid points and Reynolds numbers up to $\approx 1000$. Figure
\ref{fig:spectrum_rotstrat} shows the isotropic kinetic energy
spectrum for two $1024^3$ runs of rotating and stratified flows forced
at small scales with fixed Froude number ($\textrm{Fr} = 0.04$) and
two different Rossby numbers ($\textrm{Ro} = 0.08$ and $0.16$). Unlike
the purely stratified case, both simulations display growth of energy
at large scales, negative energy flux (not shown here), and a spectrum
compatible with a $\sim k^{-5/3}$ power law at small scales. The
simulation with $\textrm{Ro} = 0.08$ shows a larger peak of the
energy spectrum, taking place at smaller wave numbers ($k \approx
4$). As both spectra are displayed at the same time, this indicates
that the inverse cascade is faster in this simulation (which has 
$N/f = 2$), compared with the simulation with $\textrm{Ro} = 0.16$
which has $N/f = 4$.

\begin{table}
\caption{\ADDA{Parameters of the $512^3$ simulations in
  Secs.~\ref{sec:qgnumerics} and \ref{sec:spacetime}: $N$ is the
  Brunt-V\"ais\"ala frequency, $f$ is the Coriolis frequency,
  $\textrm{Fr}$ is the Froude number, $\textrm{Ro}$ is the Rossby
  number, $\epsilon$ is the energy injection rate, $k_\textrm{Oz}$ is
  the Ozmidov wavenumber, and $k_\Omega$ is the Zeman wavenumber (the
  maximum resolved wavenumber in $512^3$ runs is 
  $512/3 \approx 170$).}}
\label{table:params}
\centering
\begin{ruledtabular}
\begin{tabular}{c c c c c c c c}
Run & $N$ & $f$ & Fr & Ro &$\epsilon$&$k_\textrm{Oz}$&$k_\Omega$\\
\hline
$1$ & $0.25$ & 1 & $0.8$   & $0.2$ & $0.01$ & $1.2$ & 10 \\ 
$2$ & $0.5$   & 1 & $0.4$   & $0.2$ & $0.01$ & $3.5$ & 10 \\ 
$3$ & $2$      & 1 & $0.1$   & $0.2$ & $0.01$ & 28      & 10 \\ 
$4$ & $5$      & 1 & $0.04$ & $0.2$ & $0.01$ & 112    & 10 \\ 
$5$ & $0.5$   & 2 & $0.3$ & $0.075$ & $0.005$ & 5    & 40 \\ 
$6$ & $1$      & 1 & $0.2$   & $0.2$ & $0.02$ & 7        & 7   \\
$7$ & $4$     & 8 & $0.08$ & $0.04$ & $0.007$ & 95 & 270 \\ 
$8$ & $8$     & 4 & $0.05$ & $0.1$ & $0.003$ & 413 & 146 \\
\end{tabular}
\end{ruledtabular}
\end{table}

The results of the detailed parametric study of the effect of varying
$N/f$ on the inverse cascade is summarized in
Fig.~\ref{fig:marino}. Figure \ref{fig:marino}(a) shows the kinetic
energy as a function of time, $E_V(t)$, for several simulations with
varying $\textrm{Ro}$ and $\textrm{Fr}$ numbers. The simulations are
forced at small scales and started from random initial conditions at
wave numbers $k > k_F$ (where $k_F$ is the forcing wavenumber). First
the energy decays as the system develops a turbulent spectrum, and
after $\approx 5$ turnover times two regimes can be observed. In the
simulation without rotation ($f=0$), $E_V(t)$ remains approximately
constant. In the simulation without stratification ($N=0$), energy
grows monotonously in time. The same happens in the rotating and
stratified cases, but the two simulations with $N/f=1/2$ and $N/f=2$
display the fastest growth of the energy.

\begin{figure}
\includegraphics[width=8cm]{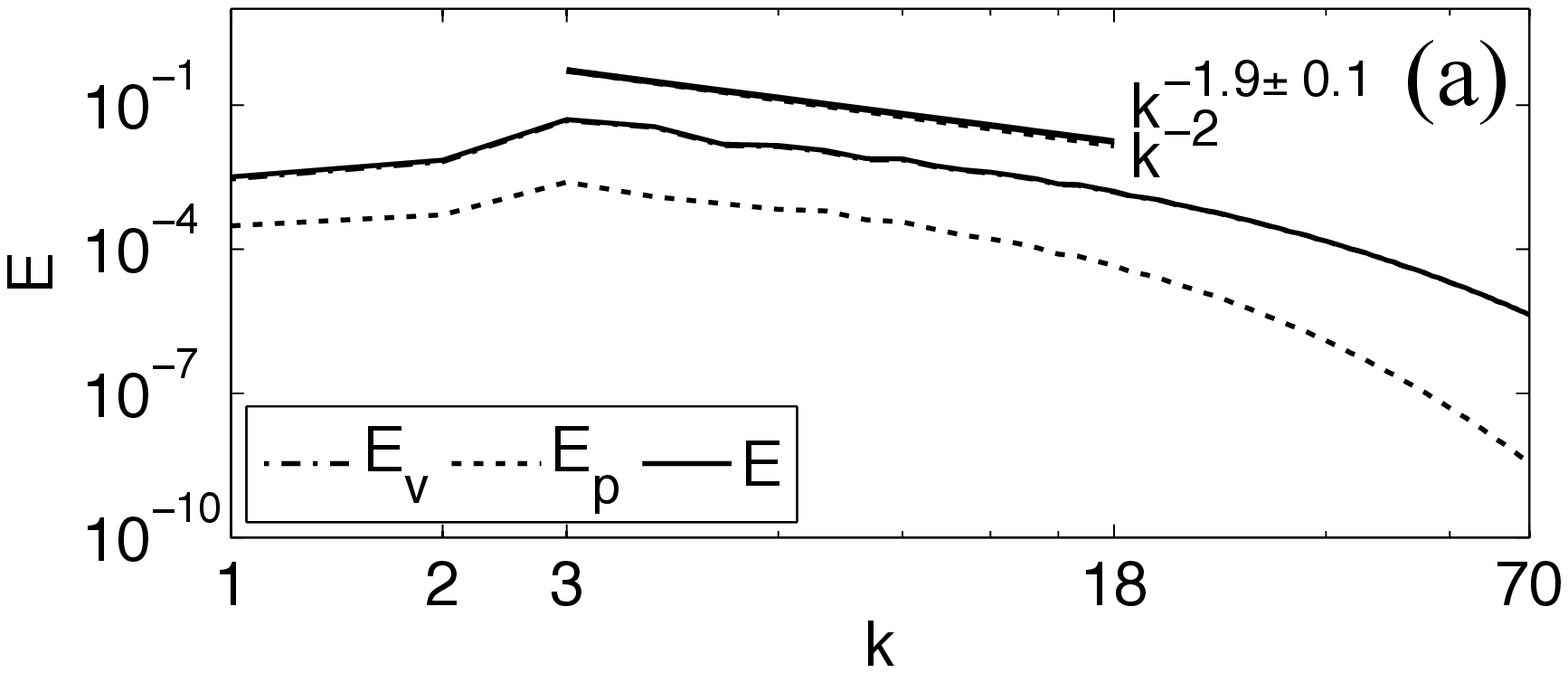}
\includegraphics[width=8cm]{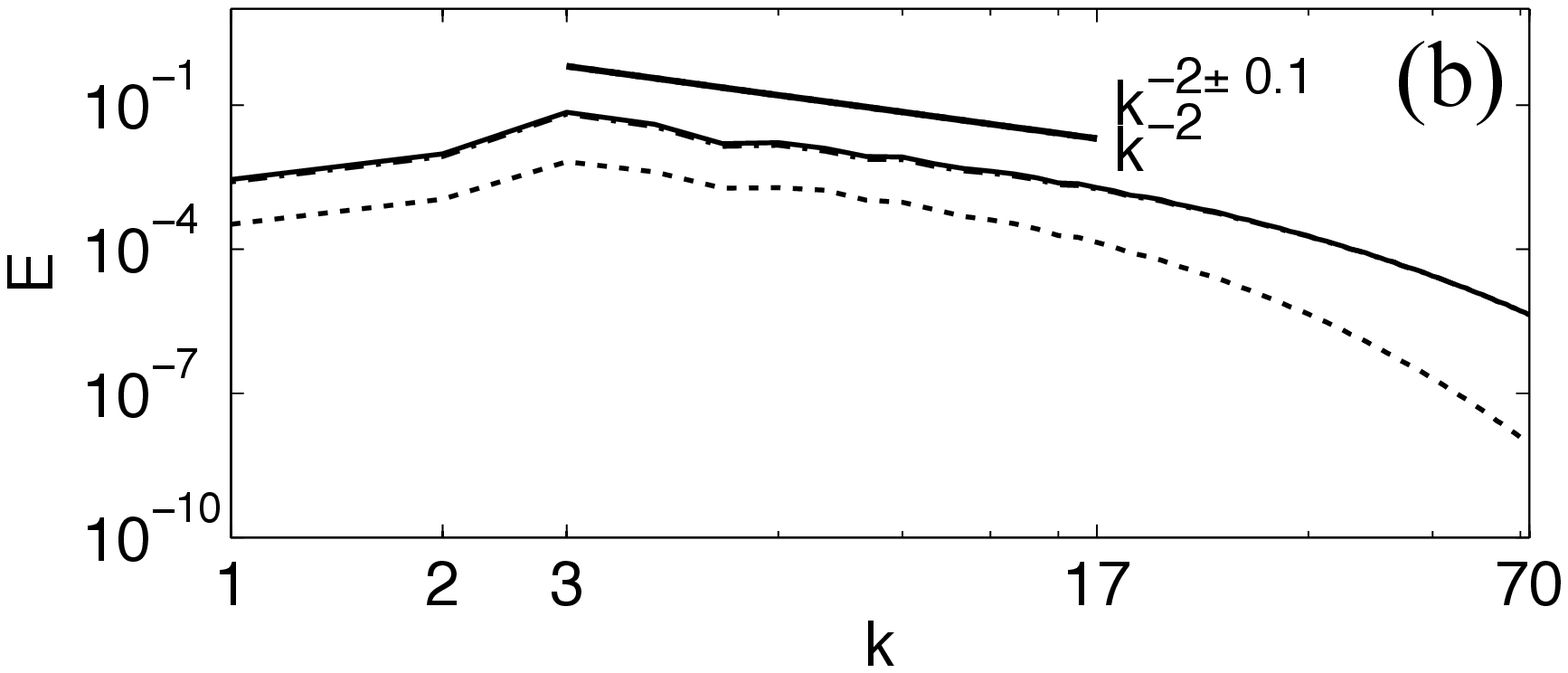}
\includegraphics[width=8cm]{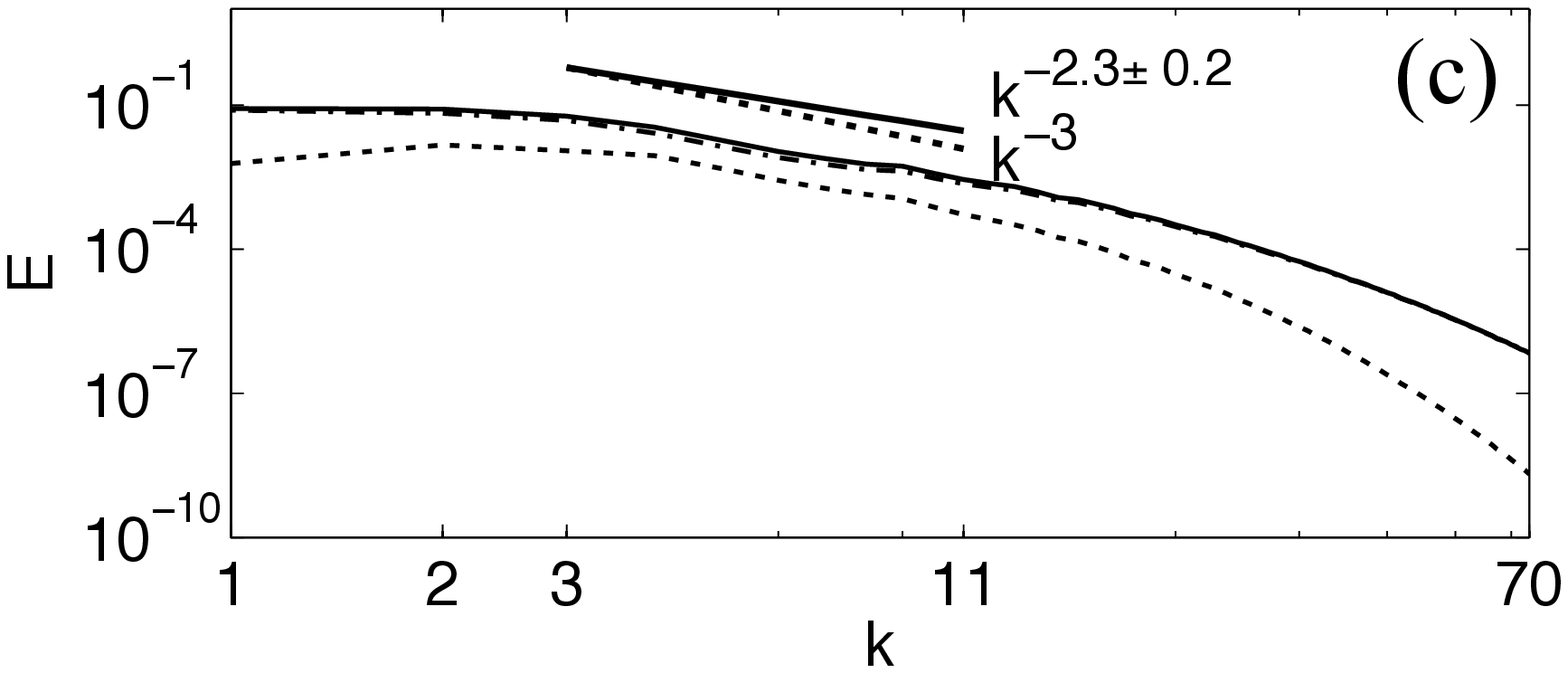}
\includegraphics[width=8cm]{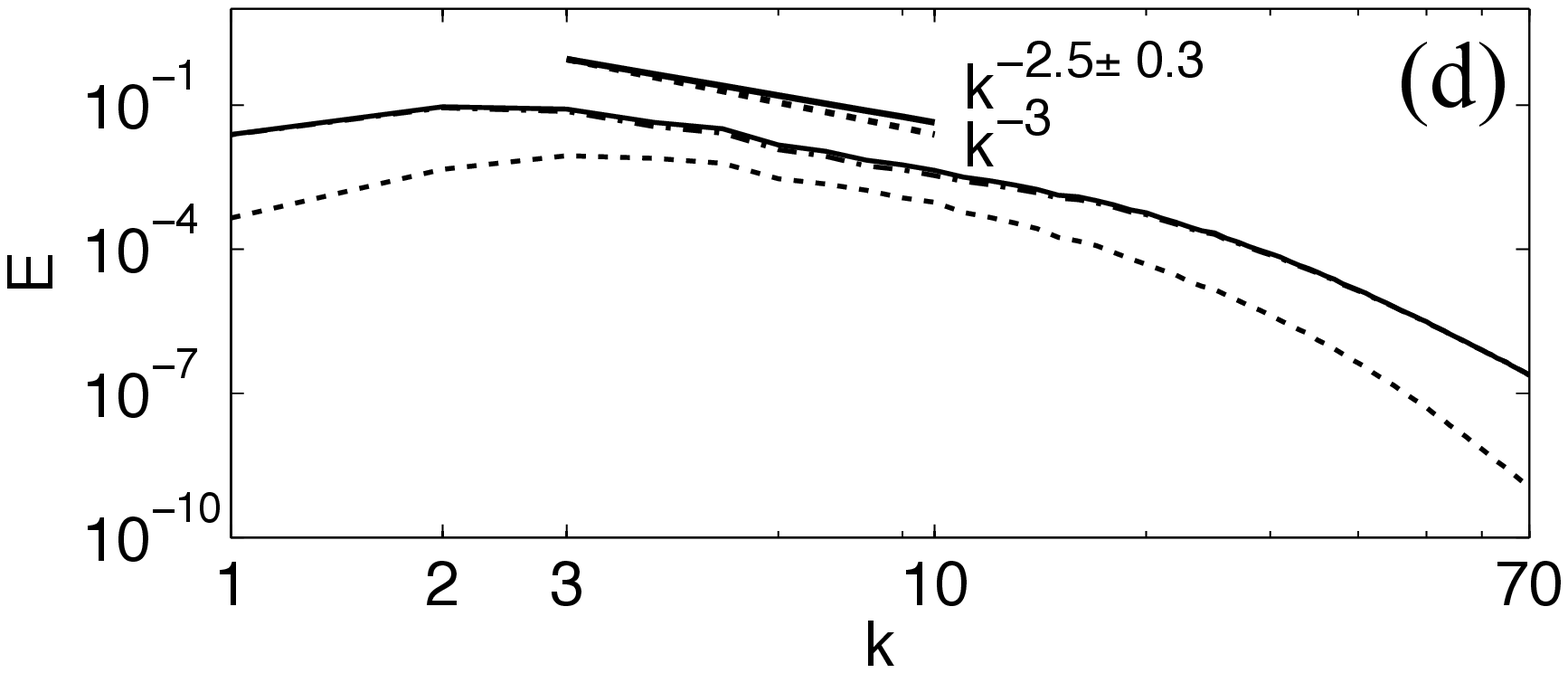}
\caption{Detail of the direct cascade inertial range in the isotropic
  spectra of kinetic energy $E_V$, power spectrum of temperature
  fluctuations $E_P$, and total energy $E$, in several $512^3$
  simulations of rotating and stratified turbulence, all forced at
  $k_F \approx 3$ and with $\textrm{Ro} \approx 0.2$. 
  (a) $\textrm{Fr} = 0.8$, $N/f = 1/4$ (run 1), 
  (b) $\textrm{Fr} = 0.4$, $N/f = 1/2$ (run 2), 
  (c) $\textrm{Fr} = 0.1$, $N/f = 2$ (run 3), and 
  (d) $\textrm{Fr} = 0.04$, $N/f = 5$ (run 4). Several power laws are
  shown as references.}
\label{fig:spectrum_iso}
\end{figure}

A parametric study of this growth speed as a function of $N/f$ for
multiple runs is shown in Fig.~\ref{fig:marino}(b). Two sets of
simulations are shown, one with fixed $\textrm{Ro}$ and varying $N/f$
by varying the $\textrm{Fr}$ number, and another with fixed
$\textrm{Fr}$ and varying $\textrm{Ro}$. In all cases, simulations
with $1/2 \le N/f \le 2$ display the fastest growth of the energy. In
\cite{Marino13} it was shown that this growth corresponds to a growth
of the energy in 2D modes at large scales, and that it is caused by an
inverse energy cascade with constant (and negative) energy flux which
also takes maximum values in the range $1/2 \le N/f \le 2$. In
\cite{Marino13} it was thus speculated that the larger efficiency of
the inverse cascade in this range was associated with the absence of
resonant interactions and the prevalence of QG behavior in this region
of parameter space. In the rest of this paper we will consider new
simulations and analysis to confirm this is indeed the case.

\section{\label{sec:qgnumerics}Quasi-geostrophic behavior}

\begin{figure}
\includegraphics[width=8cm]{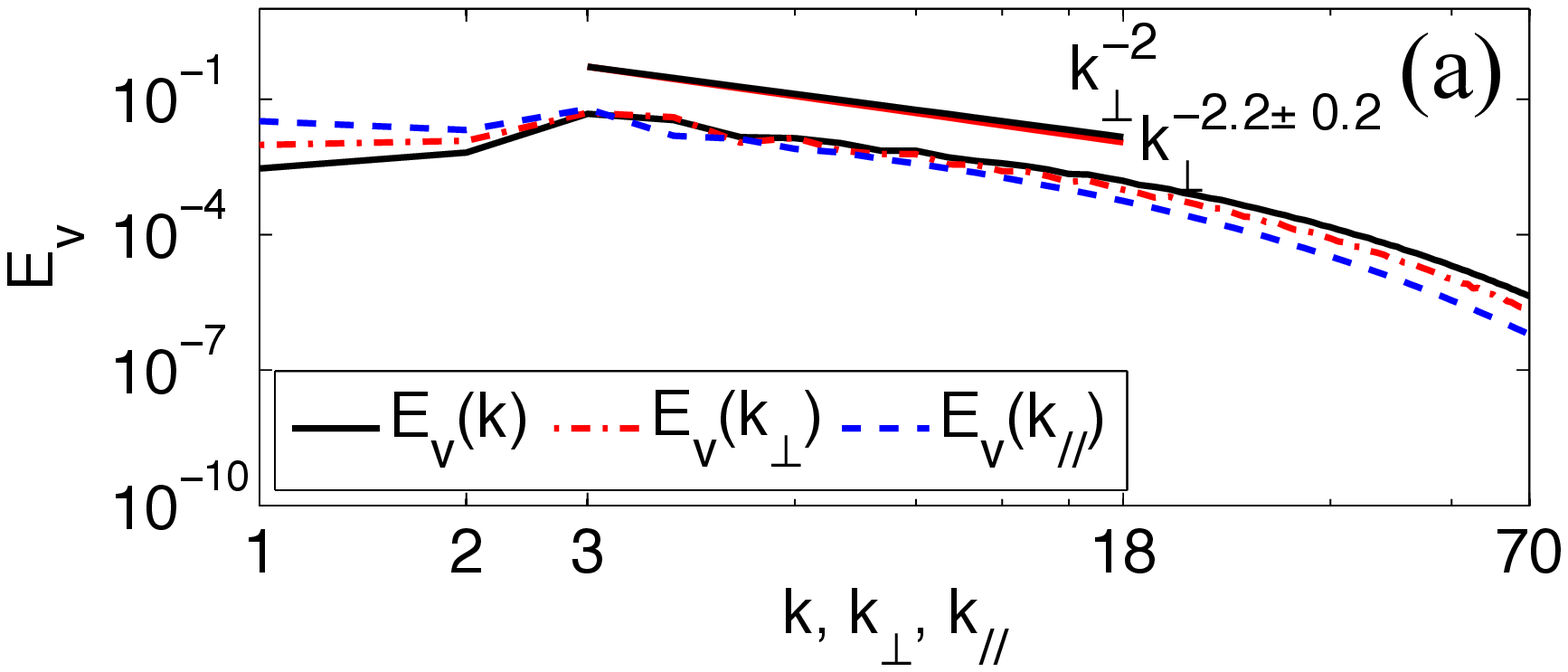}
\includegraphics[width=8cm]{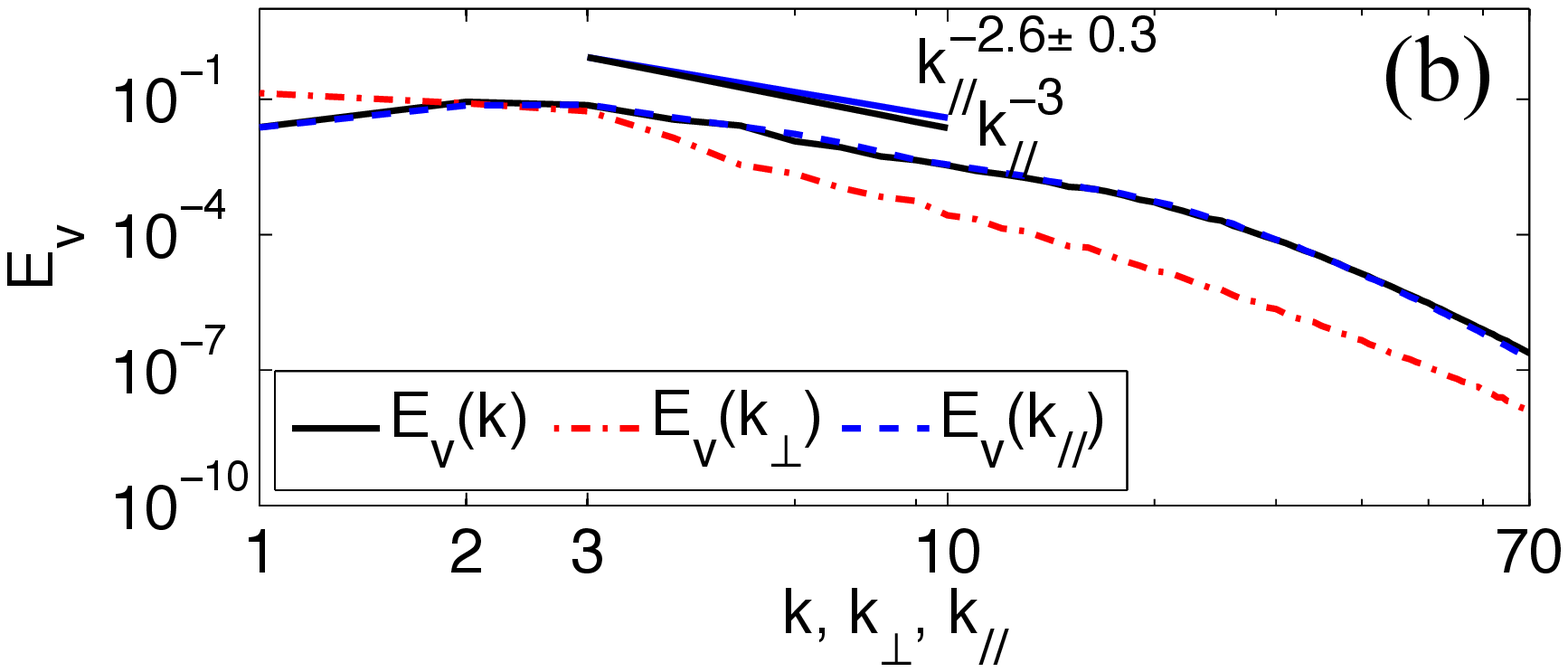}
\caption{({\it Color online}) 
  Detail of the direct cascade inertial range in the isotropic
  spectrum of kinetic energy $E_V(k)$, reduced perpendicular spectrum
  of kinetic energy $E_V(k_\perp)$, and reduced parallel spectrum of
  kinetic energy $E_V(k_\parallel)$, for two $512^3$ simulations of
  rotating and stratified turbulence forced at $k_F \approx 3$ with
  $\textrm{Ro} \approx 0.2$. 
  (a) $\textrm{Fr} = 0.8$, $N/f = 1/4$ (run 1), and
  (b) $\textrm{Fr} = 0.04$, $N/f = 5$ (run 4). Several power laws are
  shown as references.}
\label{fig:spectrum_all}
\end{figure}

For the analysis in this section and in the next, we performed a new
set of simulations of rotating and stratified turbulence. As in 
Sec.~\ref{sec:spacetime} we will perform a spatio-temporal analysis of
the data to extract the strength of the waves (which requires storage
of the data with very high cadence in time), we will only be able to
consider moderate spatial resolutions. Therefore, while simulations in
the previous sections had spatial resolutions of $1024^3$ grid points
or larger, simulations in the next two sections will consist of two
large datasets of runs with $256^3$ and $512^3$ grid points. As in the
previous sections, we will explore parameter space by considering
multiple values of $\textrm{Fr}$, and for each of these values we will
vary $\textrm{Ro}$ to explore the effect of changing $N/f$. Overall,
we performed 15 simulations with $256^3$ grid points with 
$\textrm{Re} \approx 800$, 
$0.02 \lesssim \textrm{Fr} \lesssim 1.8$ \ADDA{($0.1 \le N \le 10$)},
$0.04 \lesssim \textrm{Ro} \lesssim 0.3$ \ADDA{($0.6 \le N \le 8$)},
and $0.1 \lesssim N/f \lesssim 10$, and one simulation without
rotating nor stratification ($f=N=0$). We also performed 8 simulations
with $512^3$ grid points with $\textrm{Re} \approx 1500$, 
$0.05 \lesssim \textrm{Fr} \lesssim 0.8$, 
$0.04 \lesssim \textrm{Ro} \lesssim 0.2$, and 
$0.25 \lesssim N/f \lesssim 5$. \ADDA{The parameters and
  characteristic wave numbers for these 8 simulations are given in
  detail in Table \ref{table:params}.}

Also, as resolution in these simulations is limited, we forced the
flow at much larger scales leaving very little room for the growth of
energy at large scales. We thus use Taylor-Green forcing
\cite{Mininni08} acting at $k_F=3\sqrt{2} \approx 3$. Large scale
forcing is required as one of our goals will be to identify the role
of the waves at small scales, but it will force us to complement the
results in these sections with the results in the previous sections
where time resolution was not as good, but the inverse cascade
ranges were better resolved. To study both ranges separately is a
common practice in geophysical fluid dynamics due to constraints in
computing power, as only very recently simulations were able to
resolve dual cascades in a unique simulation  at very high resolution
\cite{Marino15}.

\begin{figure}
\includegraphics[width=4cm]{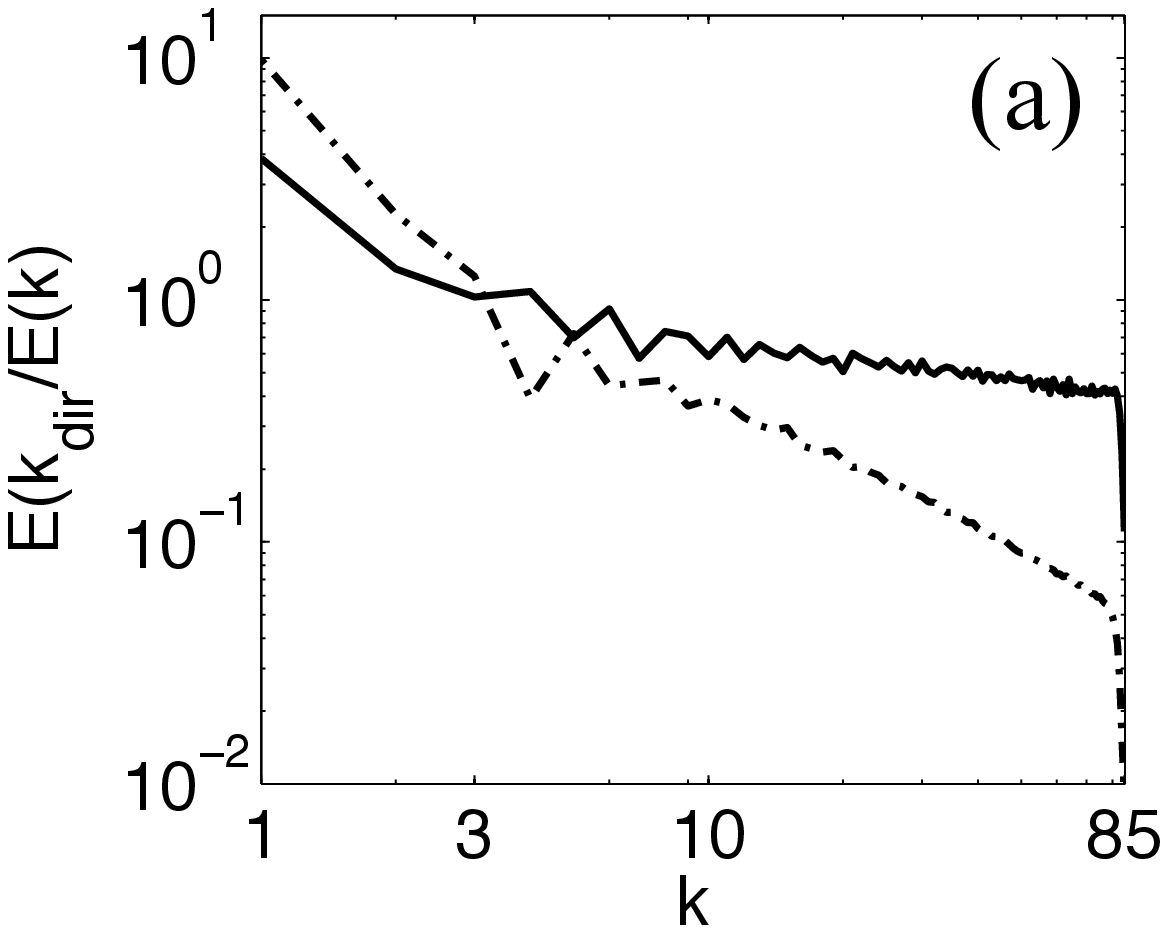}\includegraphics[width=4cm]{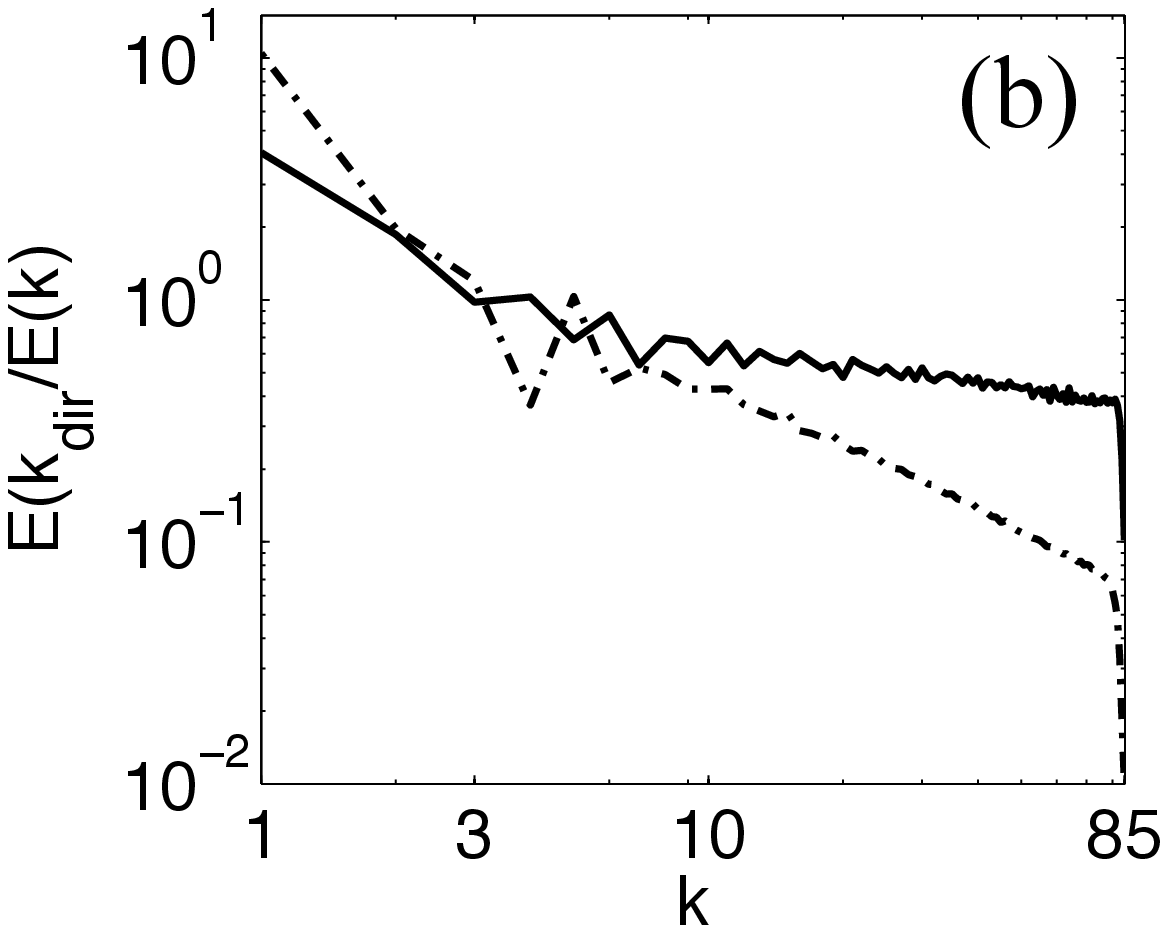}\\
\includegraphics[width=4cm]{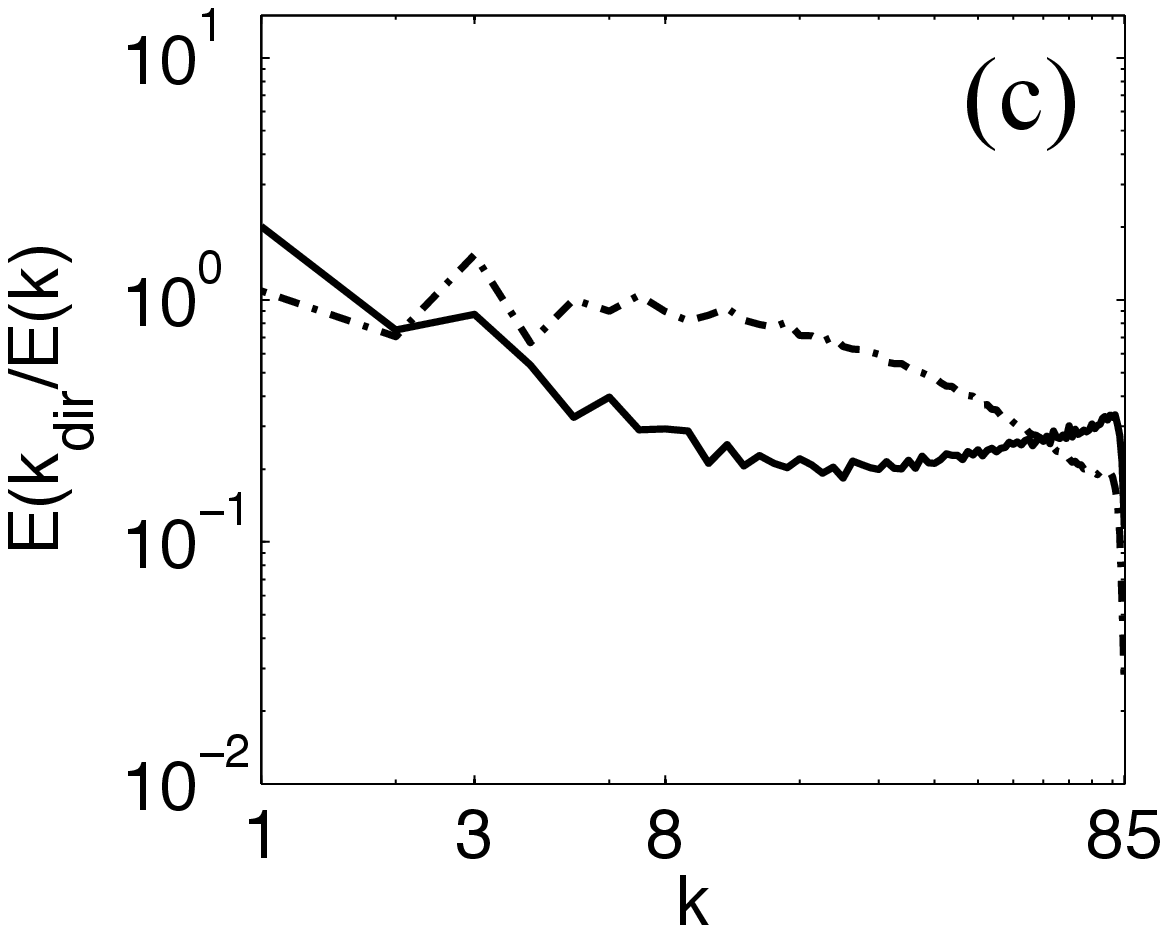}\includegraphics[width=4cm]{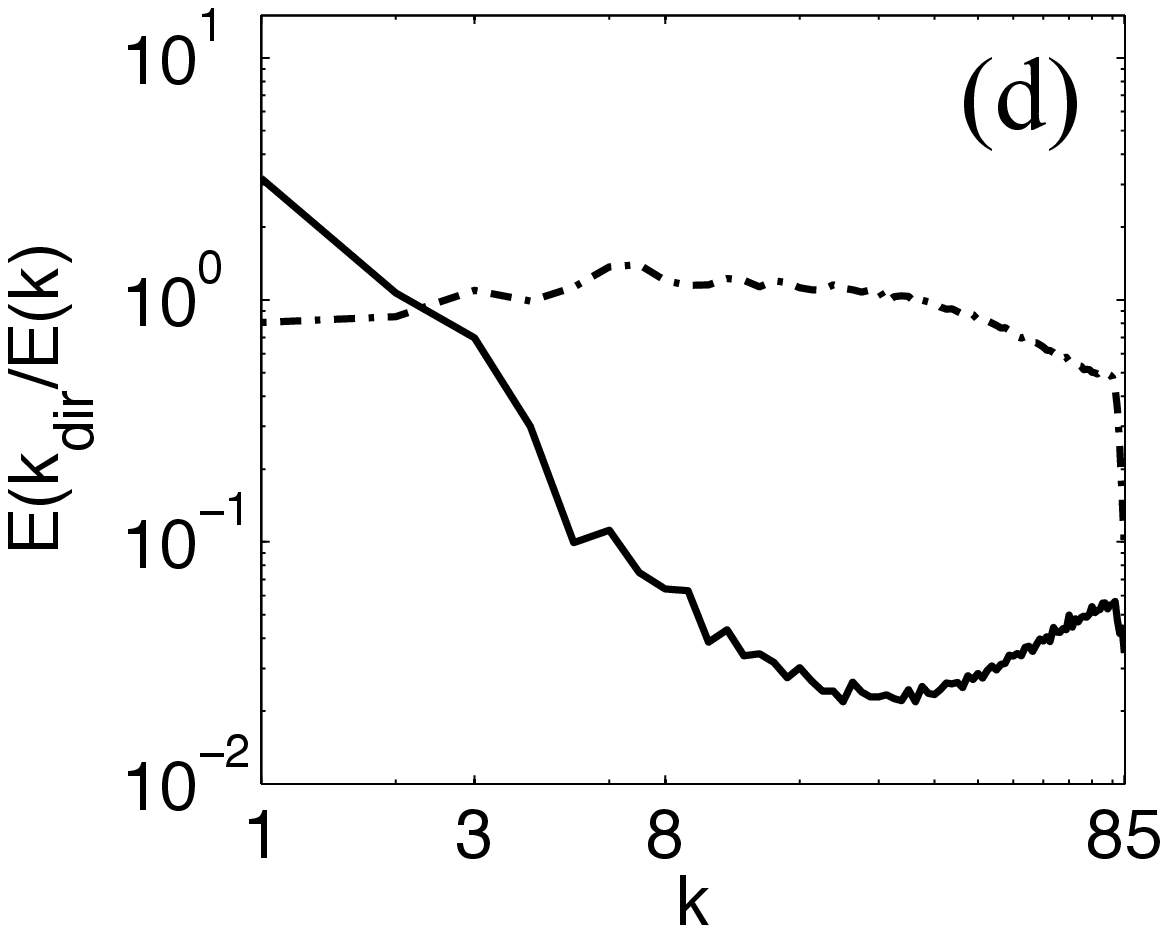}
\caption{Ratio of the spectra $E(k_\perp)/E(k)$ (solid) and
  $E(k_\parallel)/E(k)$ (dash-dotted) for several $256^3$ simulations
  of rotating and stratified turbulence forced at $k_F \approx 3$ with 
  $\textrm{Ro} \approx 0.2$. 
  (a) $\textrm{Fr} = 0.8$, $N/f = 1/4$,
  (b) $\textrm{Fr} = 0.4$, $N/f = 1/2$,
  (c) $\textrm{Fr} = 0.1$, $N/f = 2$, and
  (d) $\textrm{Fr} = 0.04$, $N/f = 5$.}
\label{fig:spectrum_ratio}
\end{figure}

\begin{figure}
\includegraphics[width=4.2cm]{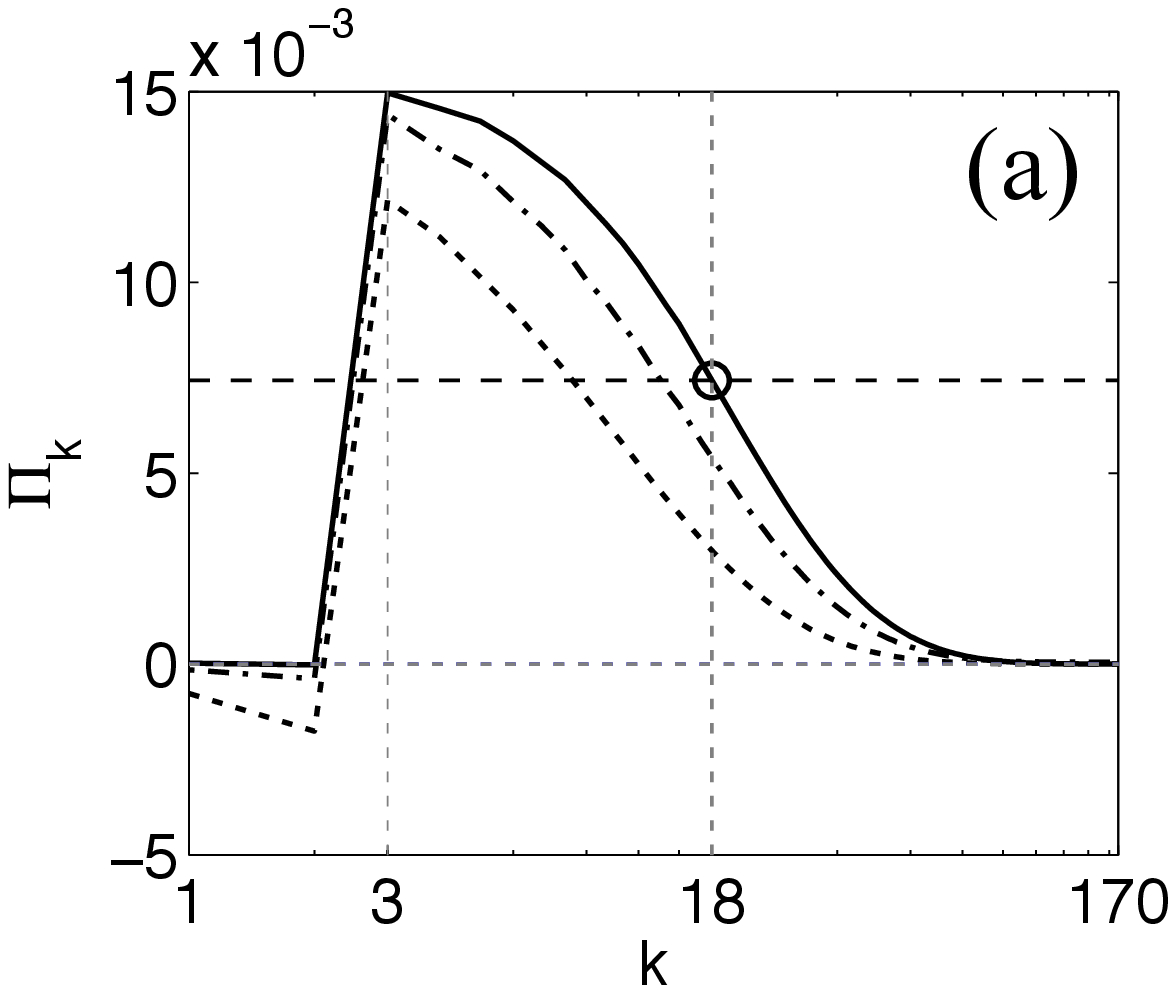}\includegraphics[width=4.2cm]{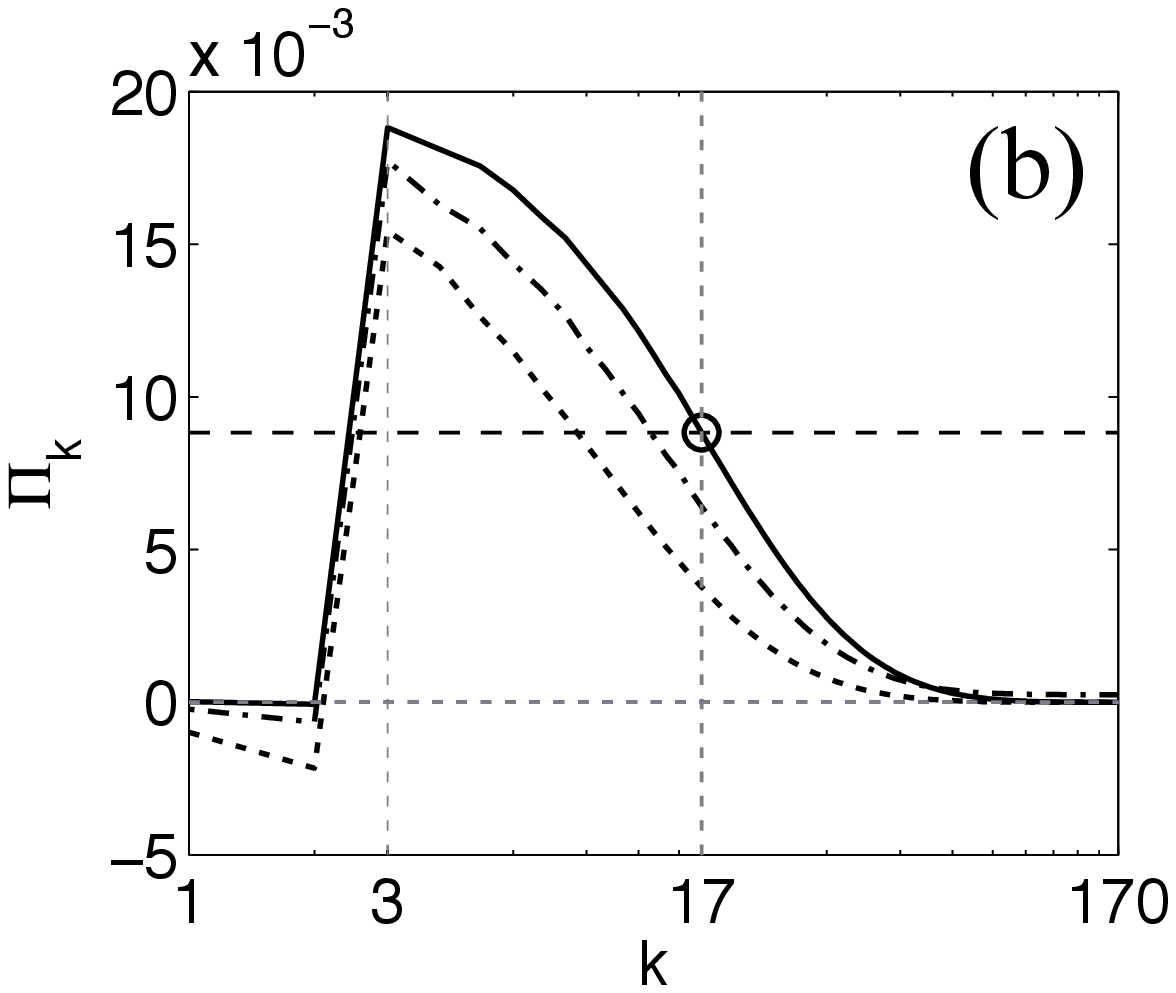}\\
\includegraphics[width=4.2cm]{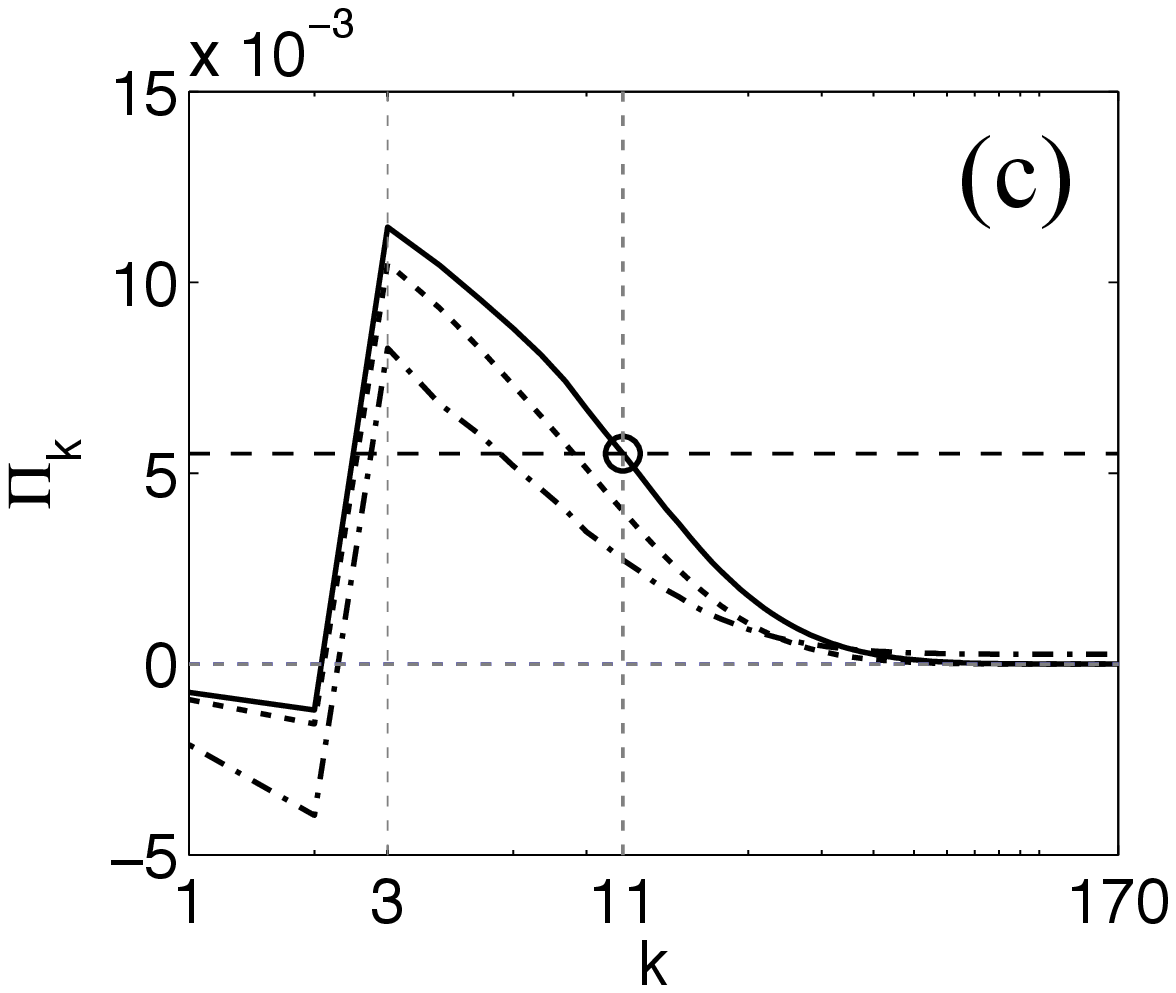}\includegraphics[width=4.2cm]{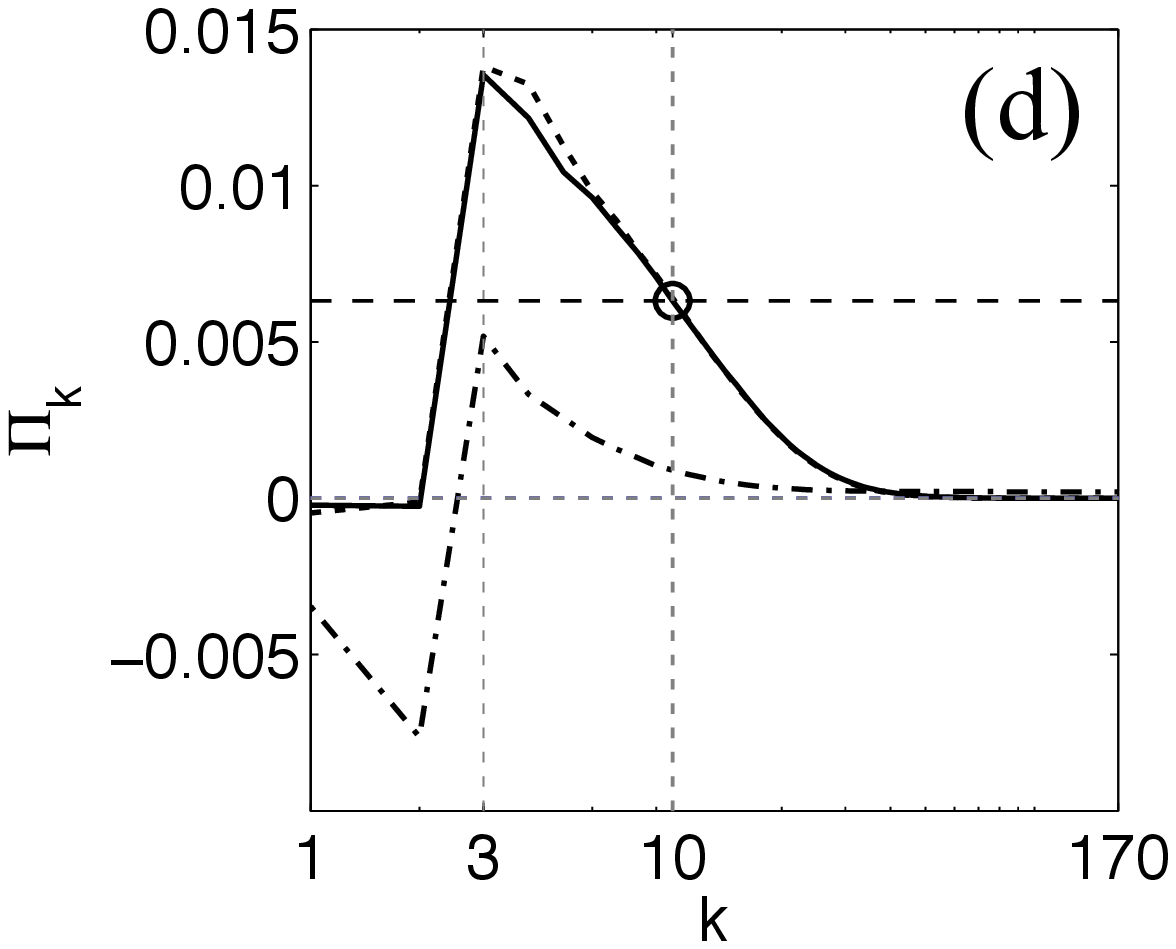}
\caption{Isotropic energy flux $\Pi(k)$ (solid), perpendicular energy
  flux $\Pi(k_\perp)$ (dashed), and parallel energy flux
  $\Pi(k_\parallel)$ (dash-dotted) in several $512^3$ simulations
  of rotating and stratified turbulence forced at $k_F = 3$ with $Ro
  \approx 0.2$. From left to right and top to bottom: 
  (a) $\textrm{Fr} = 0.8$, $N/f = 1/4$ (run 1), 
  (b) $\textrm{Fr} = 0.4$, $N/f = 1/2$ (run 2), 
  (c) $\textrm{Fr} = 0.1$, $N/f = 2$ (run 3), and
  (d) $\textrm{Fr} = 0.04$, $N/f = 5$ (run 4). The straight lines and
  the circles indicate the wave number at which the flux drops to
  $1/2$ of its value at the forcing wave number $k_F\approx 3$. Note
  that the largest negative isotropic and perpendicular fluxes (for
  $k<k_F$) are obtained in the case with $N/f = 2$. The simulation
  with $N/f = 5$ has negative parallel flux, but negligible isotropic
  and perpendicular inverse fluxes at large scales.}
\label{fig:flux}
\end{figure}

\begin{figure}
\includegraphics[width=8.5cm]{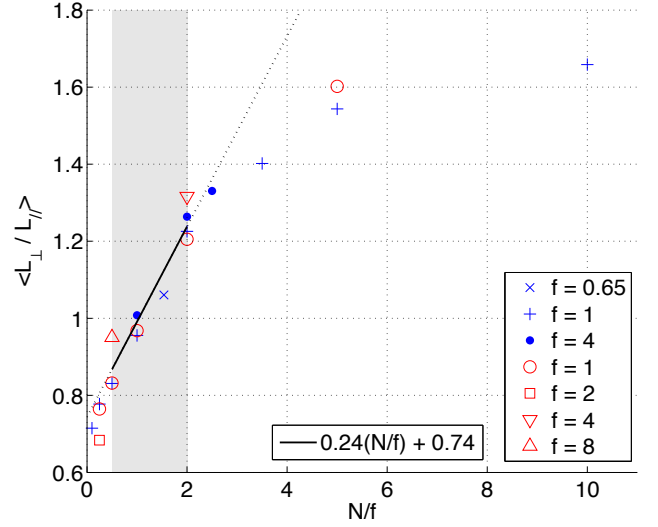}
\caption{({\it Color online})
  Ratio of the perpendicular to parallel integral scales
  $\left<L_\perp/L_\parallel\right>$ (averaged over time in the
  turbulent steady state) as a function of $N/f$ for several $256^3$
  and $512^3$ simulations of rotating and stratified turbulence with
  varying $\textrm{Ro}$ and $\textrm{Fr}$ numbers. The range with no
  resonant interactions, $1/2 \le N/f \le 2$, is indicated by the gray
  shading. In this range, a linear fit of
  $\left<L_\perp/L_\parallel\right>$ as a function of $N/f$ (expected
  from quasi-geostrophy) is shown as a reference.}
\label{fig:LLvsNf}
\end{figure}

Figure \ref{fig:spectrum_iso} shows a detail, in the vicinity of the
inertial range, of the isotropic energy spectrum $E(k)$ for four
$512^3$ runs with $\textrm{Ro} \approx 0.2$ and varying $\textrm{Fr}$
(and therefore, varying $N/f$). Power laws $\sim k^{-2}$ and $\sim
k^{-3}$ are shown as references. Also as a reference we present a
best fit to a power law in the range of scales $k \in [k_F,k_{1/2}]$,
where $k_{1/2}$ is the wave number at which the energy flux drops to
$1/2$ of its value at the injection scale, i.e., 
$\Pi(k_{1/2}) = \Pi(k_F)/2$. This fit is not intended to claim a
specific power law followed by the inertial range, as the value of
$1/2$ used to define the drop in the flux is arbitrary, but rather to
illustrate that as $N/f$ increases, the energy spectrum becomes
steeper, going from the behavior expected for a purely rotating flow
($\sim k^{-2}$) to that of a purely stratified flow ($\sim
k^{-3}$). \ADDA{This can be also understood from the values of the
  Zeman and Ozmidov wavenumbers in Table \ref{table:params}. In
  Fig.~\ref{fig:spectrum_iso}(a), which corresponds to run 1,
  $k_\textrm{Oz}\approx 1.2$ and all intermediate wave numbers are
  dominated by rotation, at least until $k_\Omega \approx 10$. As the
  Brunt-V\"ais\"ala frequency is increased, the Ozmidov wavenumber
  also increases, and more scales are dominated by both rotation and
  stratification. The energy spectrum of run 4, shown in
  Fig.~\ref{fig:spectrum_iso}(d), corresponds to $k_\textrm{Oz}\approx
  112$ and $k_\Omega \approx 10$. Thus, in this case stratification is
  dominant at all scales.} Also, note that a small growth of energy at
wave numbers $k<k_F$ can be observed in some of these runs. Although
the scale separation at large scales (small wave numbers) in these
runs is small to study the inverse cascade, inverse transfer can be
observed, as will be also shown in the energy fluxes.

\begin{figure}
\includegraphics[width=8cm]{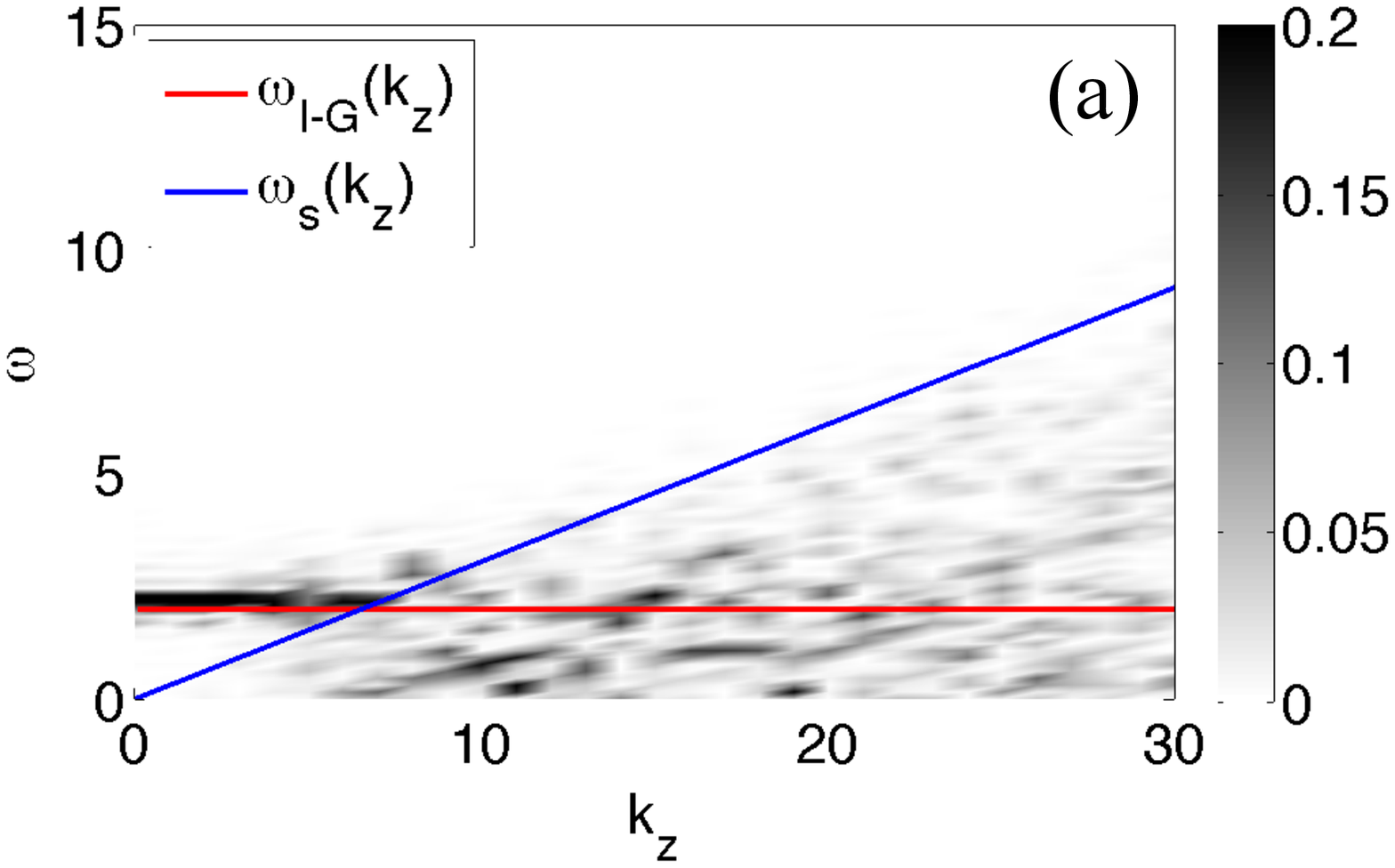}
\includegraphics[width=8cm]{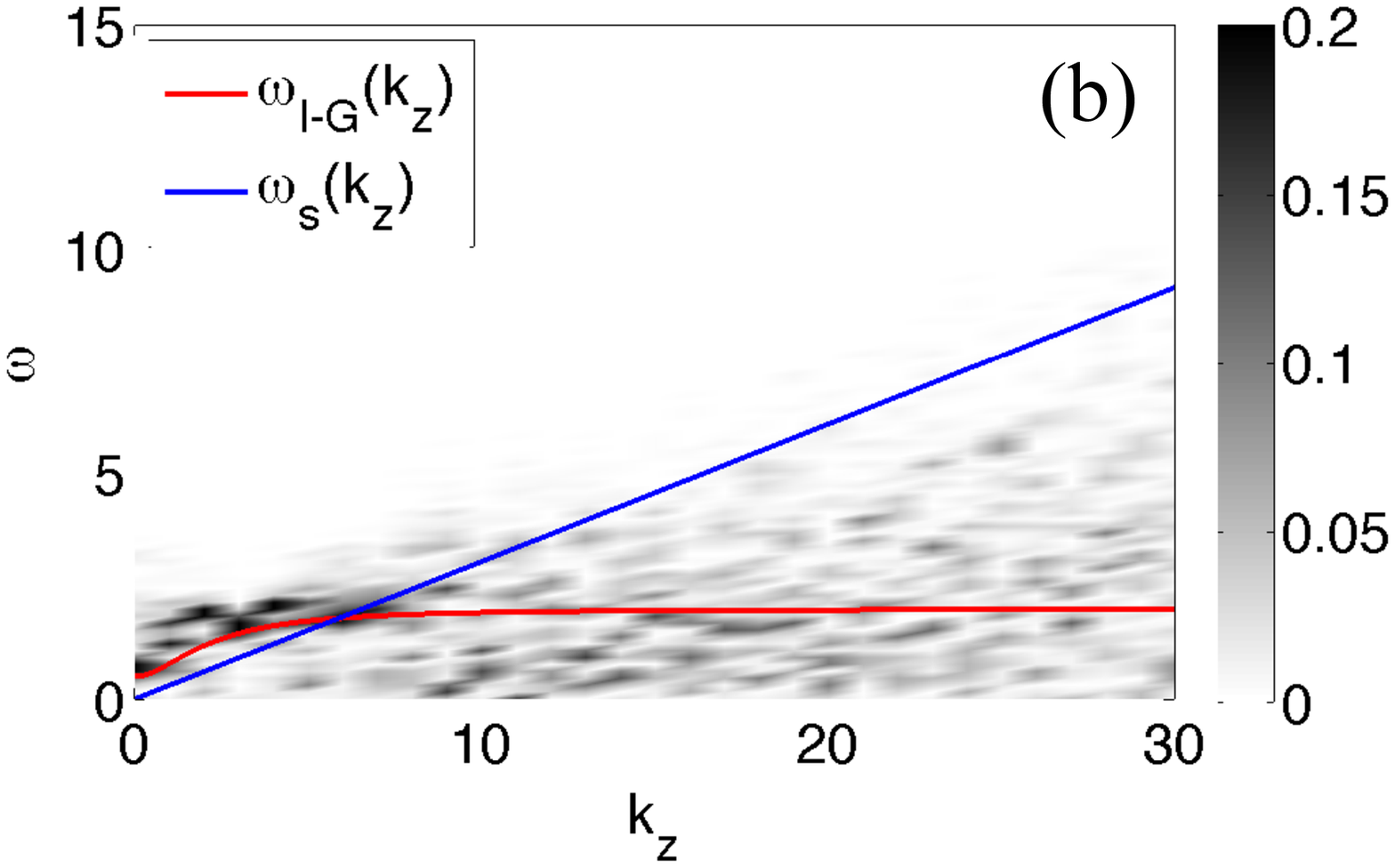}
\includegraphics[width=8cm]{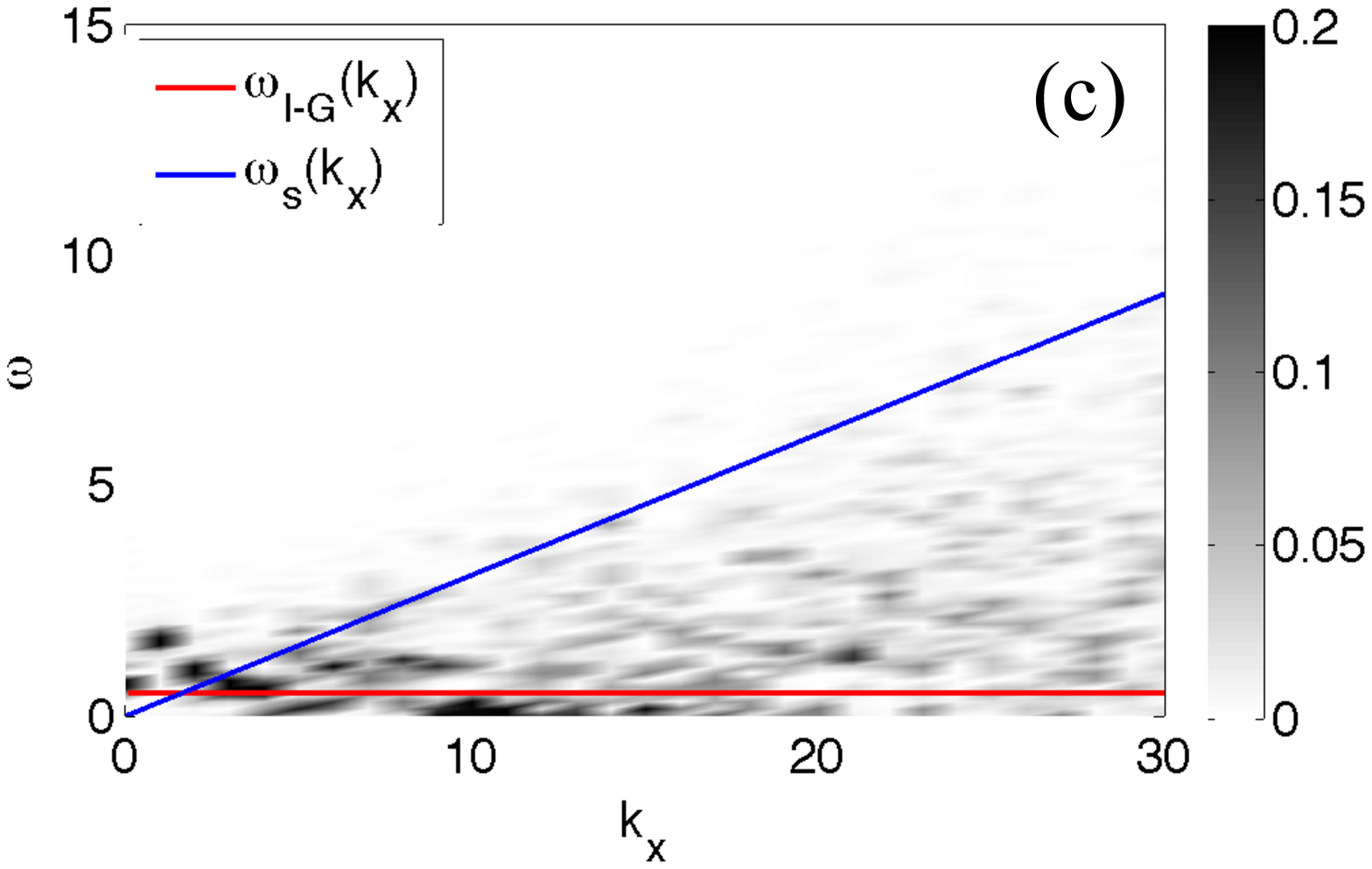}
\includegraphics[width=8cm]{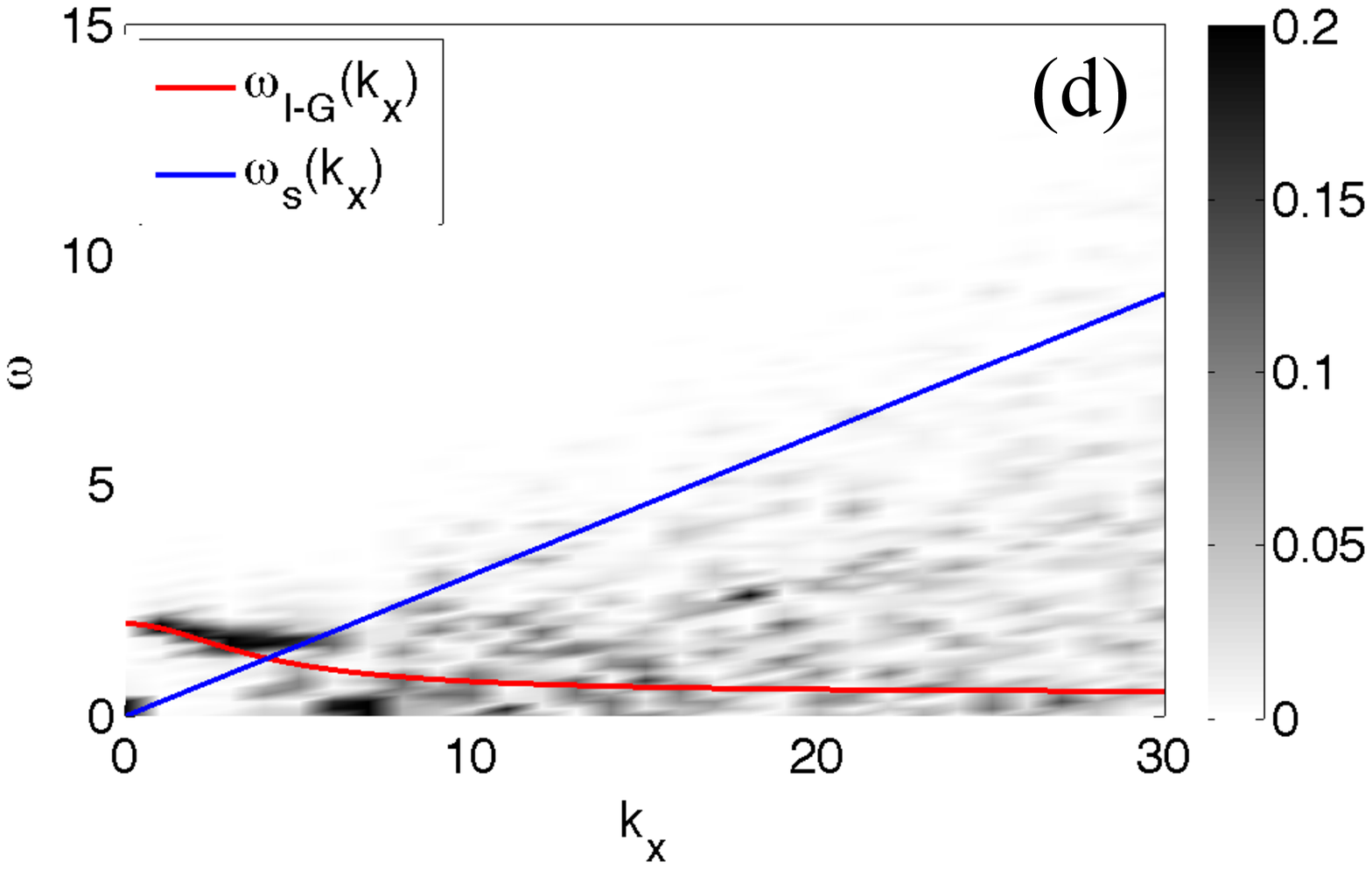}
\caption{({\it Color online})
  Spatio-temporal power spectra $|u_x|^2(k_x,k_y,k_z,\omega)$
  and $|u_z|^2(k_x,k_y,k_z,\omega)$ for a $512^3$ simulation of
  rotating and stratified turbulence with $\textrm{Fr}=0.3$, 
  $\textrm{Ro} = 0.075$ ($N/f = 1/4$, run 5 in Table
  \ref{table:params}). Two-dimensional slices of each four-dimensional
  spectrum are shown by fixing two components of the wave vector. 
  (a) $|u_x|^2(k_x=0,k_y=0,k_z,\omega)$,
  (b) $|u_x|^2(k_x=3,k_y=0,k_z,\omega)$,
  (c) $|u_z|^2(k_x,k_y=0,k_z=0,\omega)$, and
  (d) $|u_z|^2(k_x,k_y=0,k_z=3,\omega)$. The solid lines indicate the
  dispersion relation of inertia-gravity waves,
  $\omega_\textrm{I-G}({\bf k})$, and of sweeping, 
  $\omega_s ({\bf k})$.}
\label{fig:ekw1}
\end{figure}

\begin{figure}
\includegraphics[width=8cm]{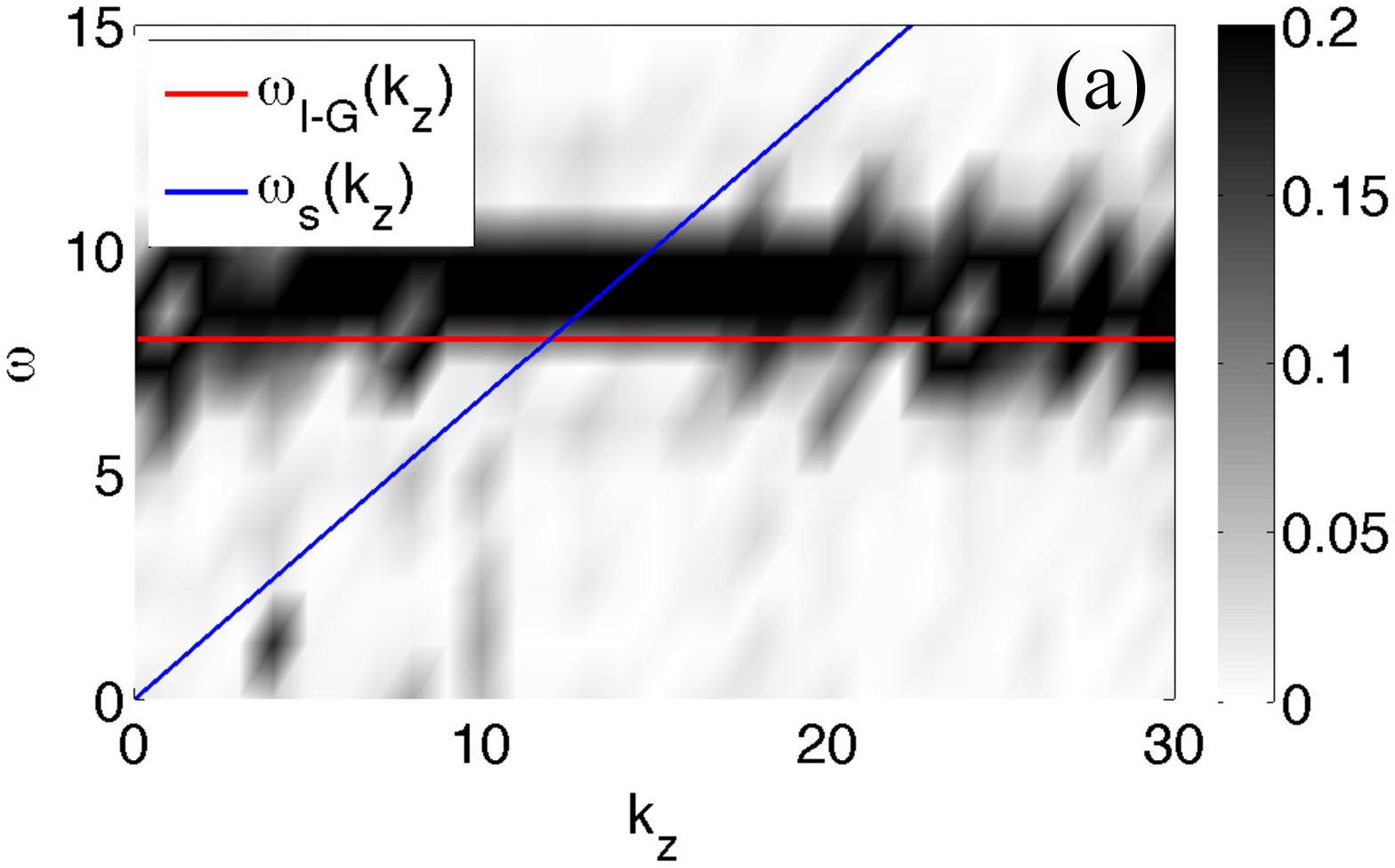}
\includegraphics[width=8cm]{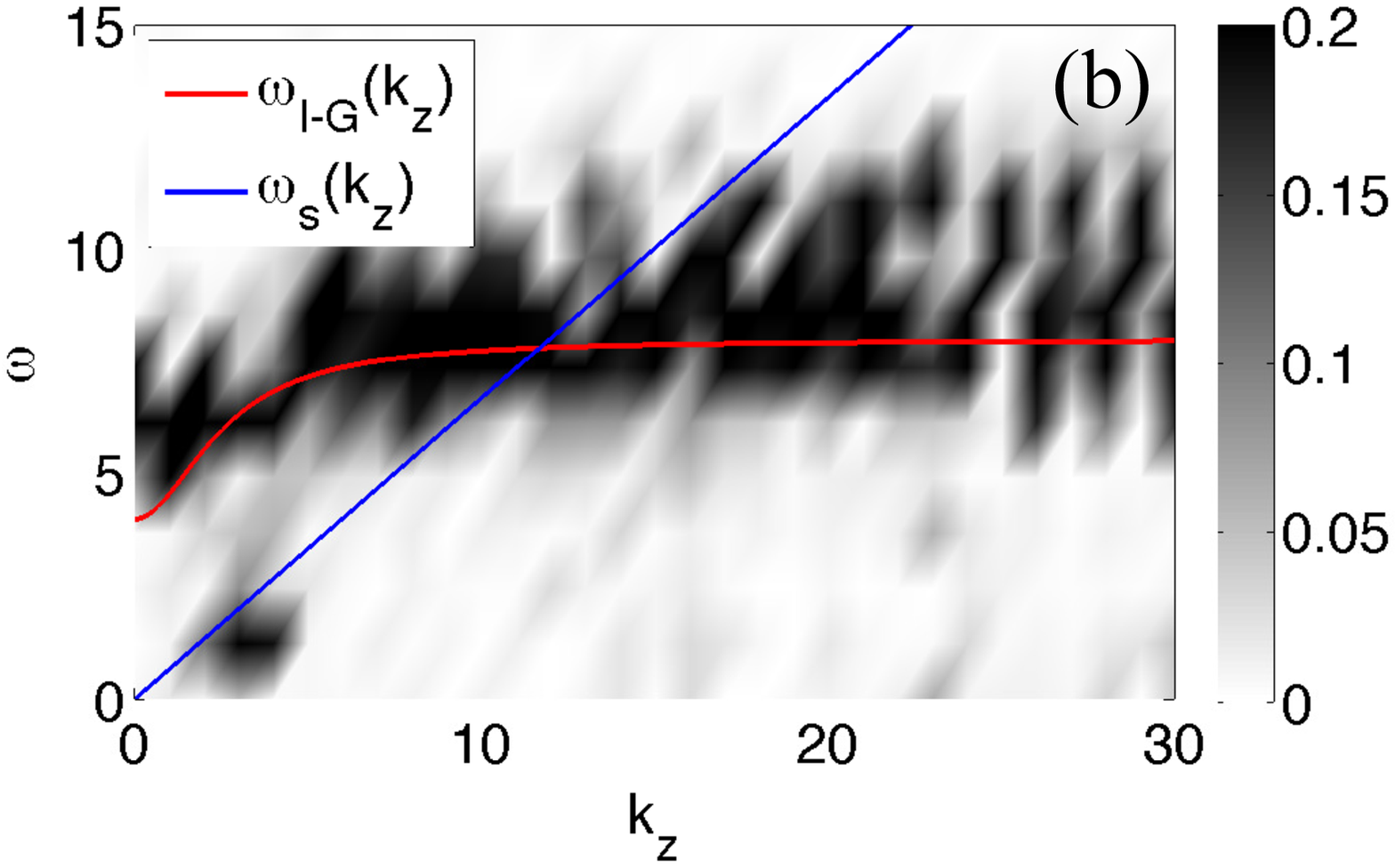}
\includegraphics[width=8cm]{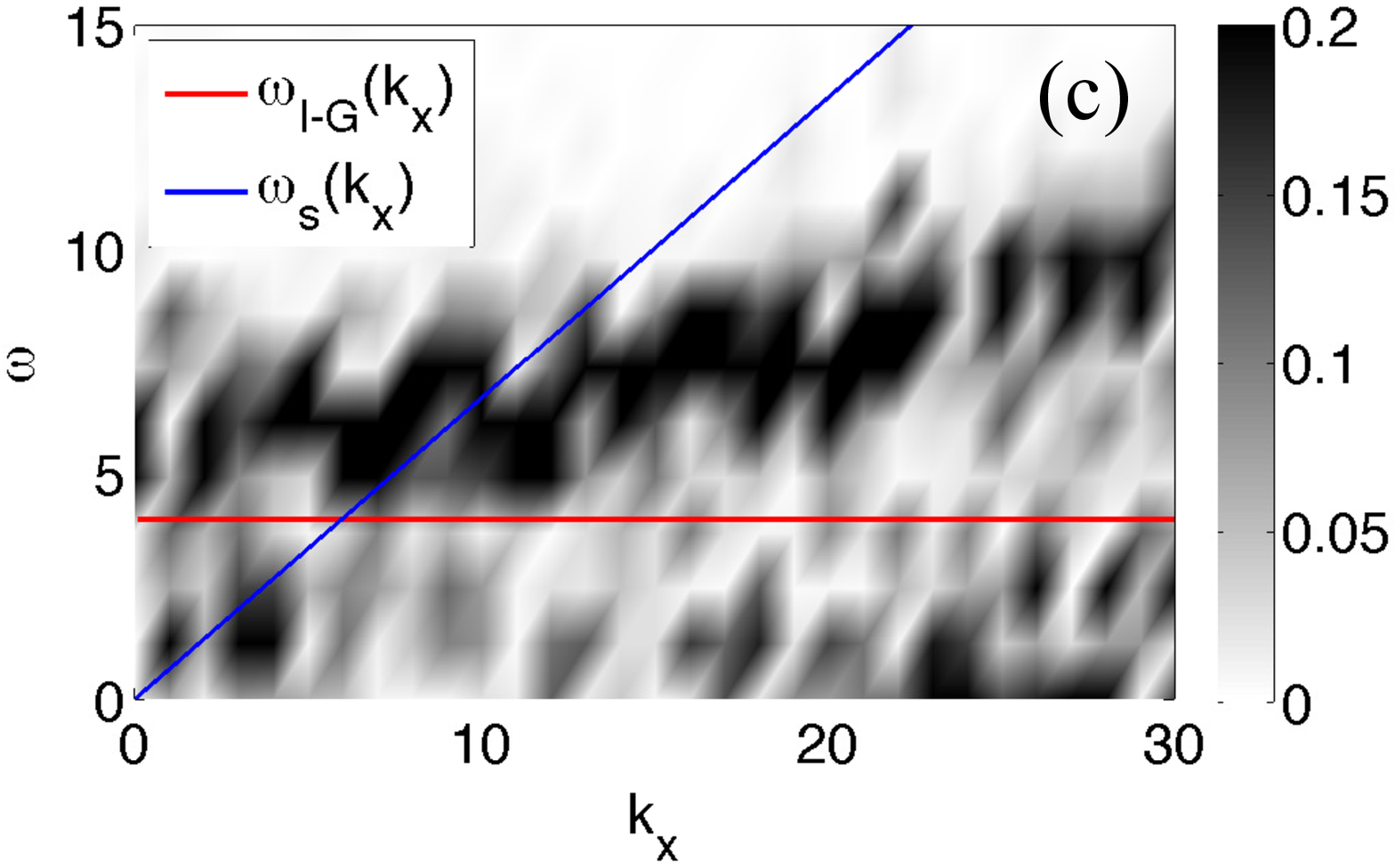}
\includegraphics[width=8cm]{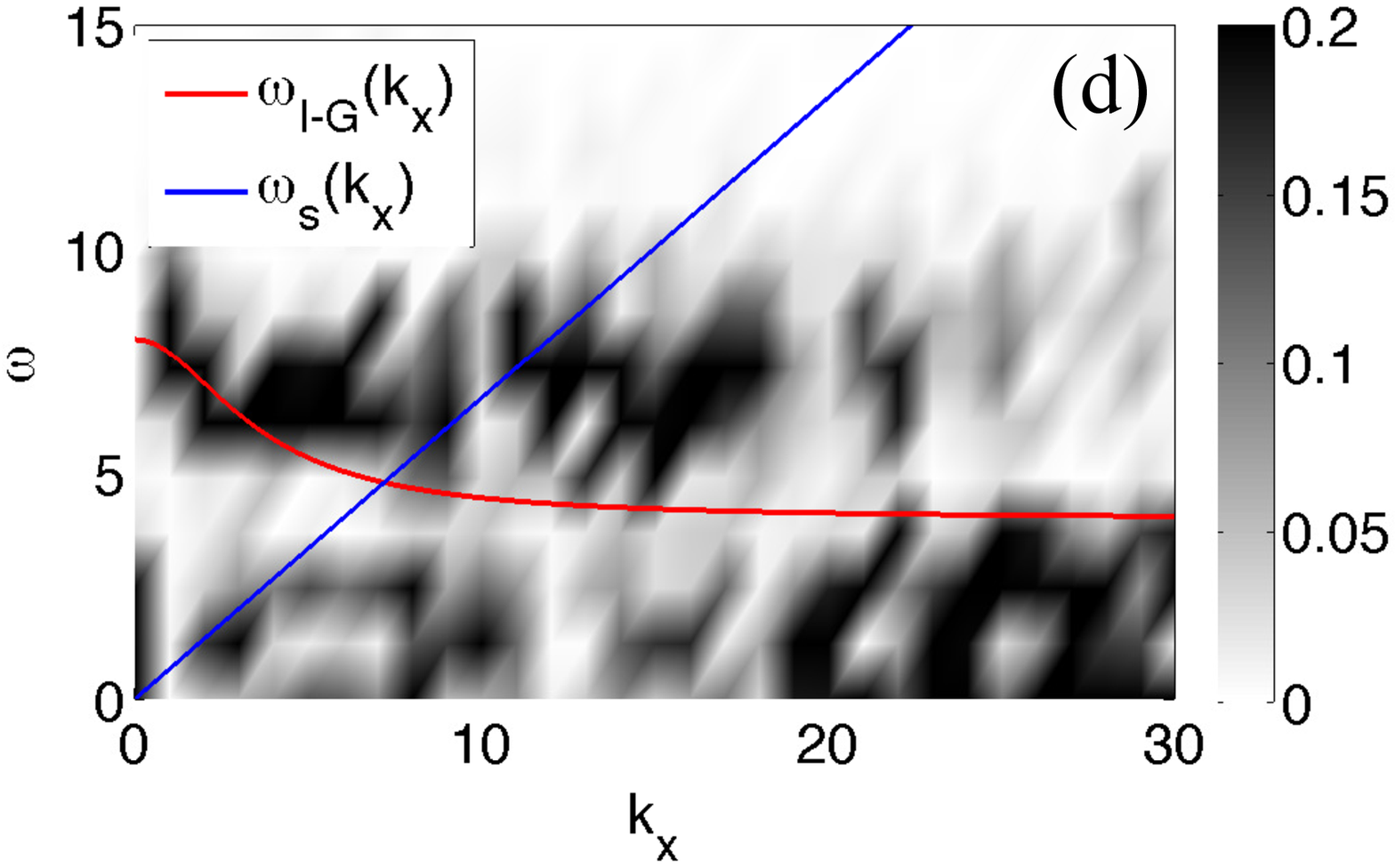}
\caption{({\it Color online})
  Spatio-temporal power spectra $|u_x|^2(k_x,k_y,k_z,\omega)$
  and $|u_z|^2(k_x,k_y,k_z,\omega)$ for a $512^3$ simulation of
  rotating and stratified turbulence with $\textrm{Fr}=0.08$, 
  $\textrm{Ro} = 0.04$ ($N/f = 1/2$, run 7 in Table
  \ref{table:params}). Two-dimensional slices of each four-dimensional
  spectrum are shown by fixing two components of the wave vector. 
  (a) $|u_x|^2(k_x=0,k_y=0,k_z,\omega)$, 
  (b) $|u_x|^2(k_x=3,k_y=0,k_z,\omega)$, 
  (c) $|u_z|^2(k_x,k_y=0,k_z=0,\omega)$, and
  (d) $|u_z|^2(k_x,k_y=0,k_z=3,\omega)$. The solid lines indicate the
  dispersion relation of inertia-gravity waves,
  $\omega_\textrm{I-G}({\bf k})$, and of sweeping, 
  $\omega_s ({\bf k})$.}
\label{fig:ekw2}
\end{figure}

Figure \ref{fig:spectrum_all} shows the isotropic and reduced parallel
and perpendicular spectra of the kinetic energy for two of the
simulations shown in Fig.~\ref{fig:spectrum_iso}. Again, power laws
and best fits to the spectrum are shown as references. While in
the inertial range for the case with $N/f = 1/4$ the reduced
perpendicular spectrum $E(k_\perp)$ is closer to the isotropic
spectrum $E(k)$, in the case with $N/f = 5$ the reduced parallel
spectrum $E(k_\parallel)$ is closer to $E(k)$. In other words, in the
case with stronger rotation most of the energy seems to accumulate
near modes with $k_\parallel \approx 0$, while in the case with
stronger stratification energy accumulates near modes with 
$k_\perp \approx 0$.

This is further illustrated in Fig.~\ref{fig:spectrum_ratio}, which
shows the ratios $E(k_\perp)/E(k)$ and $E(k_\parallel)/E(k)$ for
several simulations with varying $N/f$. For $N/f<1$ and for $k>k_F$,
the ratio $E(k_\perp)/E(k)$ is of order one in a wide range of
wave numbers, while $E(k_\parallel)/E(k)$ decreases rapidly with
increasing wave number. On the other hand, for $N/f>1$ the ratio
$E(k_\parallel)/E(k)$ remains of order one in a range of wave numbers
with $k>k_F$, while $E(k_\perp)/E(k)$ decreases rapidly. \ADDB{The
  simulations in this figure have the same Ozmidov and Zeman wave
  numbers \cite{Mininni12,Delache14,Almalkie12,Rorai15} as runs 1 to 4
  in Table \ref{table:params}. While for
  Fig.~\ref{fig:spectrum_ratio}(d) $k_\textrm{Oz} \approx 112$ and the
  Zeman wavenumber is not resolved (which is compatible with the
  larger separation between the two ratios at small scales), in all
  other cases $k_\textrm{Oz}$ and $k_\Omega$ are resolved.}

Figure \ref{fig:flux} shows the energy fluxes (isotropic, parallel, and
perpendicular) in the same four simulations as in
Fig.~\ref{fig:spectrum_iso}. Note that besides the positive flux for
$k>k_F$, there is a small backtransfer with negative flux towards
large scales for $k<k_F$. Just as in the results shown in the previous
section, with larger scale separation and a clear inverse cascade,
here the amplitude of the net negative flux is not monotonic with
$N/f$, and it becomes larger in the run with $N/f =2$. As a
comparison, the run with $N/f =1/4$ shows a smaller inverse flux, and
the run with $N/f = 5$ shows a larger inverse parallel flux but
negligible isotropic and perpendicular inverse fluxes. In the latter
case, and when larger scale separations are considered, this is just
the result of a very anisotropic flux at large scales that results in
the formation of VSHW \cite{Marino14}.

We can now use these simulations to study the role of triadic
interactions in the cascade, as well as the role of waves and
slow modes as we vary $N/f$. We start by considering anisotropy and
the scaling of typical length scales, as done before by other authors 
\cite{Charney71,Reinhaud03,Waite04,Waite06}, to then present
spatio-temporal analyses in the next section. We must then define
first parallel and perpendicular characteristic length scales, which
can be easily done from the reduced energy spectra:
\begin{eqnarray}
L_\perp &=& 2\pi \frac{\sum_{k_\perp = 1}^{k_{max}} E(k_\perp) / 
  k_\perp}{\sum_{k_\perp = 1}^{k_{max}} E(k_\perp)} , \\
L_\parallel &=& 2\pi \frac{\sum_{k_\parallel = 1}^{k_{max}} E(k_\parallel) /
  k_\parallel}{\sum_{k_\parallel = 1}^{k_{max}} E(k_\parallel)} ,
\end{eqnarray}
where these scales correspond just to an extension of the usual
isotropic integral scale to the anisotropic case. Here, $k_{max}$ is
the maximum resolved wavenumber in the simulation, which for
pseudospectral simulations using the $2/3$-rule for dealiasing
correspond to  $k_{max}=N_l/3$ with $N_l$ the linear grid resolution.

Figure \ref{fig:LLvsNf} shows the ratio of these two scales (averaged
in time) as a function of $N/f$ for several runs. In agreement with
what we observed in the anisotropic spectra
(Fig.~\ref{fig:spectrum_ratio}), the ratio 
$\left< L_\perp/L_\parallel \right>$ goes from values smaller than
one for $N/f<1$, reaches
$\left< L_\perp/L_\parallel \right> \approx 1$ for $N/f\approx 1$, and
becomes larger than one for $N/f>1$, apparently saturating for large
values of $N/f$. Moreover, this behavior is independent of the value
of $f$ (i.e., of the Froude number) as all runs seem to collapse to
the same curve. But note also that there is a range of $N/f$ for which 
$\left< L_\perp/L_\parallel \right>$ seems to scale linearly with
$N/f$, and which seems to be in agreement with the region indicated in
all previous figures corresponding to the range in which there are no
resonant interactions, $1/2 \le N/f \le 2$. In this region QG modes
are expected to dominate, whereby the linear relationship can be
expected from Charney's argument \cite{Charney71} that a turbulent QG 
flow is isotropic in the re-scaled coordinate $(N/f) z$, which implies
a linear relation for the vertical integral scale $L_\parallel \sim (f/N)
L_\perp$, or equivalently
\begin{equation}
\left< \frac{L_\perp}{L_\parallel} \right> = A \frac{N}{f} + B .
\label{eq:fit}
\end{equation}
In Fig.~\ref{fig:LLvsNf} we also show a best fit for this relation to
our data. This scaling was reported before in numerical simulations of
QG turbulence \cite{Reinhaud03}, where the authors found 
$\left< L_\perp / L_\parallel \right> \approx 1.3 (N / f)$, \ADDA{and
  also recently observed in simulations of freely decaying stratified
  turbulence \cite{Dritschel15}.} As our definitions of the vertical
and parallel length scales are not equivalent to those used in
\cite{Reinhaud03}, the prefactor cannot be directly compared. But in
both cases it is remarkable that a linear relation holds, specially in
the $1/2 \le N/f \le 2$ range in our case. \ADDB{Outside this range,
  the limits of pure rotation and pure stratification are of
  interest. For pure rotation $\left< L_\perp / L_\parallel \right>$
  is of order unity, as $L_\parallel \approx 2\pi$ as a result of the
  formation of column-like structures and as $L_\perp$ approaches
  $2\pi$ as the result of the inverse cascade of energy. For a purely
  stratified flow the ratio depends on the Froude number (or on $N$),
  in reasonable agreement with Billant and Chomaz scaling
  \cite{Billant01} which states that $L_\parallel \approx U/N$.}

\section{\label{sec:spacetime}Spatio-temporal spectrum}

\begin{figure}
\includegraphics[width=8cm]{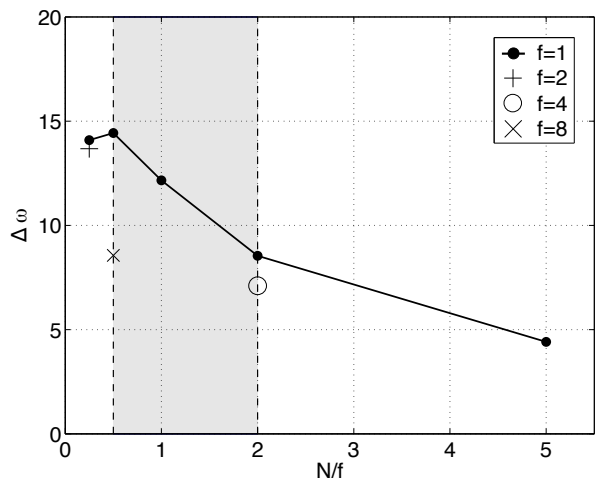}
\caption{Measured dispersion of the energy around the theoretical
  dispersion relation of inertia-gravity waves, as a function of $N/f$
  for several $256^3$ and $512^3$ simulations of rotating and
  stratified turbulence with varying $\textrm{Ro}$ and $\textrm{Fr}$
  numbers. The range with no resonant interactions, $1/2 \le N/f \le
  2$, is indicated by the gray area. The dispersion for $f=1$ takes
  its maximum in this region. A few points for other values of $f$ are
  also shown.} 
\label{fig:dispersion}
\end{figure}

In this section we present the spatio-temporal spectrum of several of
the simulations discussed in Sec.~\ref{sec:qgnumerics}. While
extraction of waves and slow modes is often done using normal mode
decompositions of the frozen in time fields in Fourier space
\cite{Metais96,Herbert14,Marino15b}, \ADDA{these approaches are based
  on linearized equations and thus can mix ageostrophic modes with
  balanced components of the flow. Better identification of balanced
  modes can be achieved using higher-order expansions of the equations
  \cite{Dritschel15}.} 
However, a precise separation of waves, mean
winds, and eddies requiere information resolved in space and in
time. In \cite{Clark15b} we introduced the spatio-temporal spectrum as
a way to do this decomposition, and showed multiple applications
including purely rotating \cite{Clark14b} and purely stratified flows 
\cite{Clark15}. \ADDB{The spatio-temporal spectrum was also used in
  recent laboratory experiments of rotating turbulence to study
  inertial waves \cite{Yarom14,Campagne15}.} The main objective of
this section is to consider the rotating and stratified case, in
particular in the range $1/2 \le N/f \le 2$, and to use the
spatio-temporal spectrum to quantify the relevance of wave modes and
of slow modes. As discussed in Secs.~\ref{sec:inverse} and
\ref{sec:qgnumerics}, several effects in this range are believed to be
associated with a dominance of slow modes and a relatively less
importance of wave modes. The spatio-temporal spectrum will allow us
to explicitly verify this is the case. However, as computation of the
spatio-temporal spectrum requires very high temporal cadence, we will
be able to do this analysis only in the simulations with $512^3$ grid
points.

\ADDA{Computation of the spatio-temporal spectrum is performed by
  storing the Fourier coefficients of the fields ${\bf u}_{\bf k}(t)$
  and $\theta_{\bf k}(t)$  with high temporal cadence (at least twice
  the period of the fastest waves). For each Fourier mode ${\bf k}$,
  the Fourier transform in time of these quantities results in
  $\hat{\bf u}({\bf k},\omega)$ and $\hat{\theta}({\bf k},\omega)$,
  which measure the phase and amplitude of each  
  $({\bf k},\omega)$-mode in a four-dimensional space. The
  spatio-temporal spectrum can then be computed, e.g., for the kinetic
  energy, as
\begin{equation}
E_V({\bf k}, \omega) = \frac{1}{2} |\hat{\bf u}({\bf k},\omega)|^2 .
\end{equation}
  This spectrum quantifies the power in each wave vector ${\bf k}$
  and frequency $\omega$, where ${\bf k}$ and $\omega$ are
  independent. In practice, waves, eddies, and other flow features
  satisfy some known relation $\omega = \omega({\bf k})$, and thus
  accumulation of energy over certain regions in the four-dimensional
  spectral space can be used to quantify how much energy is associated
  with these features. As an example, inertia-gravity waves correspond
  to an accumulation of energy in $({\bf k},\omega)$-modes satisfying
\begin{equation}
\omega = \omega_\textrm{I-G}({\bf k}),
\end{equation}
  where $\omega_\textrm{I-G}$ is the dispersion relation given in
  Eq.~(\ref{eq:omegaig}). From Eq.~(\ref{eq:sweeping}), advection of
  small-scale eddies by the large-scale velocity $U$ (i.e., sweeping)
  corresponds to the accumulation of energy in 
  $({\bf k},\omega)$-modes satisfying
\begin{equation}
\omega \lesssim k U = \omega_s({\bf k}) ,
\end{equation}
 as at each ${\bf k}$ all eddies larger than the eddy size 
 $\sim 1/k$ contribute to the Eulerian random sweeping. Doppler
 shift by a mean wind ${\bf W}$ appears as a shift of these relations
 by ${\bf W}\cdot{\bf k}$ \cite{Clark15}. A more detailed description
 of detection of energetic features in a flow using the
 spatio-temporal spectrum can be found in \cite{Clark15b}.}

Detection of waves in the four-dimensional spatio-temporal
spectrum can be also simplified with some knowledge of what components
of the fields are affected by the waves, depending on their
polarization. For systems with $N/f \ll 1$, rotation is dominant and
inertia-gravity waves reduce to inertial waves. Thus, in this case we
can look at the spectrum of $u_x$ as for inertial waves the
perturbation takes place in the plane perpendicular to
$\mbox{\boldmath $\Omega$}$. In contrast, for strongly stratified
systems with $N / f \gg 1$ we should consider $u_z$ as this is the
component of the velocity that is coupled to the temperature
fluctuations in internal gravity waves. Thus, we consider the power
spectrum of both components to allow identification of the waves as
$N/f$ is varied.

\begin{figure}
\includegraphics[width=8.5cm]{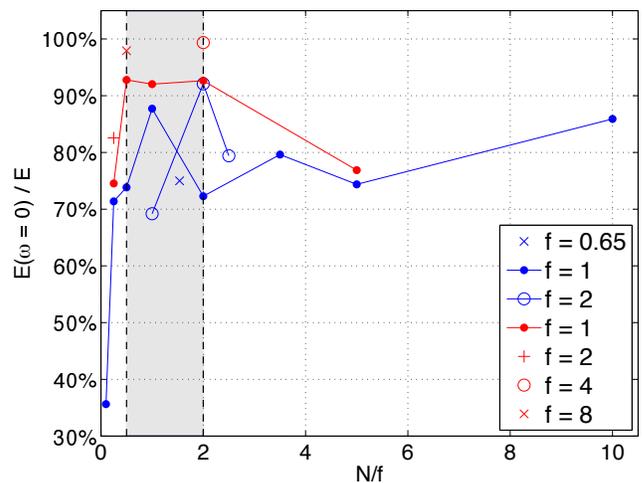}
\caption{({\it Color online}) 
  Ratio of energy in modes with zero frequency to the total
  energy as a function of $N/f$ for several $256^3$ and $512^3$
  simulations of rotating and stratified turbulence with varying
  $\textrm{Ro}$ and $\textrm{Fr}$ numbers. The range with no resonant
  interactions, $1/2 \le N/f \le 2$, is indicated by the gray
  area. Note that the ratio takes the largest values in this region.}
\label{fig:ratio_qgenergy}
\end{figure}

The spatio-temporal spectrum was computed for all $512^3$ simulations. 
Figures \ref{fig:ekw1} and \ref{fig:ekw2} show an illustration of
these spectra for the runs with $N/f=1/4$ and with $N/f = 1/2$ (this
latter run in the range $1/2 \le N/f \le 2$). \ADDA{These correspond
  respectively to runs 5 and 7 in Table \ref{table:params}.} As the
spectrum is four-dimensional, for $u_x$ we show slices for fixed
values of $k_x$ and $k_y$, and show the spectrum as a function of
$k_z$ and $\omega$, while for $u_z$ we show slices for fixed values of
$k_y$ and $k_z$, and show the spectrum as a function of $k_x$ and
$\omega$. The figures also indicate as a reference the theoretical
dispersion relation of inertia-gravity waves, and the sweeping
relation. \ADDA{The range of wave numbers shown in the figures
  correspond in all cases to wave numbers that should be dominated by
  waves, as in both simulations at least $k_\textrm{Oz}$ or $k_\Omega$
  is larger than the maximum wave number considered. In run 5
  (Fig.~\ref{fig:ekw1}) $k_\textrm{Oz} \approx 5$ and 
  $k_\Omega \approx 40$, thus rotation and stratification effects
  coexist for the smallest wave numbers, while for wave numbers in the
  range $5 \lesssim k \lesssim 40$ rotation should be more
  relevant. In run 7 (Fig.~\ref{fig:ekw2}) $k_\textrm{Oz} \approx 95$
  and $k_\Omega \approx 270$ and inertia-gravity waves can be expected
  to be dominant at all wave numbers.}

As a rule, excitation of modes lying over the theoretical dispersion
relation can be observed for wave numbers such that the frequency of
the waves is larger than the frequency of sweeping (i.e., for modes
for which the waves are faster than the sweeping). This is to be
expected as in wave turbulence, the fastest time scale controls the
decorrelation of the modes \cite{Clark14b}. But more interestingly,
the dispersion of the energy around the dispersion relation of the
waves varies with $N/f$. While the simulation with $N/f=1/4$ shows
(for small wave numbers) a sharp concentration of energy around the
relation $\omega_\textrm{I-G}({\bf k})$ (i.e., the theoretical dispersion
relation of inertia-gravity waves), for $N/f=1/2$ the dispersion of
the energy around this dispersion relation is much larger, and in some
of the figures energy is concentrated in modes that do not correspond
to wave excitations. \ADDA{Leaving aside the excitation of these
  modes associated with turbulence, the broadening of the energy near
  the dispersion relation in Fig.~\ref{fig:ekw2} can have multiple
  origins. As already discussed, sweeping results in broadening but it
  becomes dominant for wave numbers such that 
  $\omega_s(k) > \omega_\textrm{I-G}({\bf k})$. Near-resonant and
  non-resonant wave interactions, which are expected to become
  relevant for $1/2 \le N/f \le 2$, also result in broadening of the
  spectral peaks. Finally, eddy damping and turbulent fluctuations
  generate spectral broadening. Indeed, in this latter case weak
  turbulence theories and two point closures indicate that the
  broadening $\Delta \omega(k)$ is proportional to the inverse of the
  non-linear coupling time \cite{Sagaut,Nazarenko,Miquel11}. For
  strong turbulence this is the eddy turnover time $\tau_{nl}$ given
  by Eq.~(\ref{eq:nltime}). For an energy spectrum $E(k)$ between 
  $\sim k^{-3}$ and $\sim k^{-2}$, this results in $\Delta \omega(k)$
  being independent of the wave number or growing slowly as
  $k^{1/2}$.}

A detailed analysis of the spatio-temporal spectra for all simulations
is summarized in Figs.~\ref{fig:dispersion} and
\ref{fig:ratio_qgenergy}. Figure \ref{fig:dispersion} shows the net
dispersion of the energy around the theoretical wave dispersion
relation as a function of $N/f$ for all $512^3$ runs. The net
dispersion, e.g., for the power spectrum of the $x$ component of the
velocity, is computed as
\begin{equation}
\Delta \omega = \frac{\sum_{{\bf k},\omega} |\omega({\bf k}) -
  \omega_\textrm{I-G}({\bf k})|^2 |u_x({\bf k},\omega)|^2}
  {\sum_{{\bf k},\omega} |u_x({\bf k},\omega)|^2} ,
\end{equation}
and it corresponds to the the mean square differences between
$\omega_\textrm{I-G}({\bf k})$ and the actual data, weighted by the
spectral amplitude squared. As expected from the theoretical
arguments, the maximum dispersion takes place in the range 
$1/2 \le N/f \le 2$, for which resonant interactions vanish and either
strong turbulence or near-resonant or non-resonant interactions
prevail, also confirming the arguments in the previous sections. A few
points for simulations with different values of $f$ are shown, which
(for fixed $N/f$) show a smaller dispersion as $f$ is increased , to
be expected as dispersion should decrease (i.e., more energy should be
in the wave modes) as the strength of rotation and stratification
increases.

From this data we can also compute the amount of energy in slow modes,
that is to say, in modes with zero frequency. This is shown in 
Fig.~\ref{fig:ratio_qgenergy}, normalized by the total energy, for all
simulations. In the figure there is a growth in this ratio in the
region of non-resonant triads. Increasing the value of $f$ (and thus
decreasing the Froude number) seems to increase the ratio, thus
augmenting the amount of energy in modes with zero frequency. These
results are consistent with those reported in \cite{Waite06}, where
Waite and Bartello observed, for $\textrm{Ro} \approx 10^{-1}$, a
growth in vortical energy as $\textrm{Ro}$ decreases with fixed
$N$. With three different values of $N$ (4, 8 and 16), they also
concluded that the ratio of vortical energy to total energy is
independent of the Rossby number. Our data, however, shows a
dependence on the value of $\textrm{Ro}$, with smaller $\textrm{Ro}$
resulting in more energy in the slow modes. \ADDA{The results are also
  consistent with recent studies of freely decaying stratified
  turbulence with varying Prandtl ratio \cite{Dritschel15}, where it
  was found that balance prevails in the flows for $N/f \gtrsim 1$.} 
\ADDB{Finally, the trend of increasing energy in slow modes with
  decreasing Rossby is compatible with observations of inverse
  cascades in laboratory experiments of rotating flows
  \cite{Campagne14} and with numerical studies in the low Rossby
  number limit \cite{Alexakis15}, while they seem to be incompatible
  with the expected decoupling of the slow modes in that limit
  \cite{Greenspan,Cambon04}. This can be the result of finite-domain
  effects in the numerical simulations \cite{Cambon04}, or the result
  of higher-order corrections to the linear stability  theory
  \cite{Alexakis15}.}

\section{\label{sec:conclusions}Conclusions}

Inverse cascades play a central role in geophysical turbulence,
providing a mechanism for the formation of large-scale structures and
for the self-organization of disorganized flows. Since Kraichnan
contribution, extensions of his ideas to quasi-geostrophic turbulence,
rotating flows, and rotating and stratified flows have been developed
as a way to better describe atmospheric and oceanic processes. In this
context, here we reviewed several recent studies of inverse cascades
in rotating and stratified turbulence.

Special emphasis was put on reviewing the dependence of scaling laws,
anisotropy, and the strength of the inverse cascade on \ADDA{the
  inverse Prandtl ratio $N/f$} of the Coriolis frequency to the
Brunt-V\"ais\"ala frequency. We showed that while the anisotropy
and the ratio of perpendicular to parallel length scales varies
linearly with $N/f$ (in particular in the range $1/2 \le N/f \le 2$),
the strength of the inverse cascade depends non-monotonically on this
parameter, with the inverse cascade being faster in this range, and
then decreasing monotonically as $N/f$ is increased.

This behavior can be explained by considering that in the range 
$1/2 \le N/f \le 2$ no resonant triadic interactions between waves
are available. Thus, quasi-geostrophic motions are expected to
dominate the dynamics. Previous studies (see, e.g., 
\cite{Bartello95,Metais96,Reinhaud03,Waite04,Waite06,Kurien08})
have confirmed this by indirect measures, such as, e.g., normal mode
decomposition of frozen in time fields to separate wave modes from
slow modes, \ADDA{or higher-order expansions of the equations to
  separate the flow in terms of balanced and imbalanced components
  \cite{Dritschel15}.} Here we presented a new analysis based on fully
resolved spatio-temporal information, using the spectrum as a function
of the wave vector and frequency to measure how much energy is excited
in modes compatible with the dispersion relation of the waves, and in
the rest of the modes.

The simulations confirmed the linear scaling of the ratio of the
perpendicular to parallel velocity with $N/f$ in the range 
$1/2 \le N/f \le 2$, and also showed that accumulation of energy near
wave modes in the spatio temporal spectrum is minimal in this range,
while energy in slow modes becomes larger. This is in good agreement
with previous results using different methods for the analysis, and
sheds new light on reasons for the fast inverse cascade of 3D rotating
and stratified turbulent flows reported previously in \cite{Marino13}.

Fifty years after the paper by Kraichnan on inverse cascades 
in two dimensional flows \cite{Kraichnan67}, there remains much
to be done to understand turbulent phenomena in the real world. The
study of inverse cascades in more realistic scenarios has just
started, and the previous works reviewed here, as well as the new
results, consider small Reynolds numbers, periodic boundary
conditions, or infinite domains which put them far away from real
applications. As in the paper written by Kraichnan and Montgomery
\cite{Kraichnan80}, it is however our hope that some of the results of
these studies can be translated to atmospheric and oceanographic
problems. Recent studies considering observations in the atmosphere
and the ocean (see, e.g., \cite{Scott05,Sukoriansky07,Schlosser07} and
references therein) seem to indicate that indeed the gap between
idealized simulations and real measurements of the inverse cascade can
be bridged.

\begin{acknowledgments}
D.O. and P.D.M. acknowledge support from UBACYT Grant
No.~20020130100738BA, and PICT  Grants Nos.~2011-1529 and 
2015-3530. P.D.M. also acknowledges support from the CISL visitor
program at NCAR. RM acknowledges financial support from the program
PALSE ({\it Programme Avenir Lyon Saint-Etienne}) of the University
of Lyon, in the frame of the program {\it Investissements d’Avenir}
(ANR-11-IDEX-0007). AP acknowledges support form LASP, and in
particular from Bob Ergun.
\end{acknowledgments}

\bibliography{ms}

\end{document}